\documentstyle[emulateapj,apjfonts]{article}
\newcommand{\preprint}[1]{#1}

\newcommand{\url}[1]{{\tt #1}}

\slugcomment{Accepted for January 2000 ApJ Supplement}
\lefthead{White, Becker, et al.}
\righthead{The FIRST Bright Quasar Survey}

\begin{document}

\title{The {\sl FIRST} Bright Quasar Survey.  II. 60 Nights and 1200 Spectra Later}

\author{Richard~L.~White\altaffilmark{1,2},
Robert~H.~Becker\altaffilmark{2,3,4},
Michael~D.~Gregg\altaffilmark{3,4},
Sally~A.~Laurent-Muehleisen\altaffilmark{2,3,4},
Michael~S.~Brotherton\altaffilmark{2,4},
Chris~D.~Impey\altaffilmark{5},
Catherine~E.~Petry\altaffilmark{5},
Craig~B.~Foltz\altaffilmark{5},
Frederic~H.~Chaffee\altaffilmark{6},
Gordon~T.~Richards\altaffilmark{7},
William~R.~Oegerle\altaffilmark{8},
David~J.~Helfand\altaffilmark{2,9,10},
Richard~G.~McMahon\altaffilmark{10} \&
Juan~E.~Cabanela\altaffilmark{11}}
\authoremail{rlw@stsci.edu}

\altaffiltext{1}{Space Telescope Science Institute,
3700 San Martin Dr., Baltimore, MD 21218, rlw@stsci.edu}
\altaffiltext{2}{Visiting Astronomer, Kitt Peak National
Observatory, National Optical Astronomy Observatory}
\altaffiltext{3}{Physics Dept., University of California--Davis}
\altaffiltext{4}{IGPP/Lawrence Livermore National Laboratory}
\altaffiltext{5}{Steward Observatory, U. Arizona}
\altaffiltext{6}{W. M. Keck Observatory}
\altaffiltext{7}{Astronomy \& Astrophysics Center, U. Chicago}
\altaffiltext{8}{Dept. of Physics \& Astronomy, Johns Hopkins University}
\altaffiltext{9}{Astronomy Dept., Columbia University}
\altaffiltext{10}{Institute of Astronomy, Cambridge University}
\altaffiltext{11}{Dept. of Astronomy, U. Minnesota}

\begin{abstract}

We have used the VLA FIRST survey and the APM catalog of the POSS-I
plates as the basis for constructing a new radio-selected sample of optically
bright quasars.
This is the first radio-selected sample that is competitive in size
with current optically selected quasar surveys.  Using only two basic
criteria, radio-optical positional coincidence and optical morphology,
quasars and BL~Lacs can be identified with 60\% selection efficiency;
the efficiency increases to 70\% for objects fainter than
magnitude 17.  We show that a more sophisticated selection scheme can
predict with better than 85\% reliability which candidates will turn
out to be quasars.

This paper presents the second installment of the FIRST Bright Quasar
Survey with a catalog of 636 quasars distributed over 2682 square
degrees.  The quasar sample is characterized and all spectra are
displayed.  The FBQS detects both radio-loud and radio-quiet quasars
out to a redshift $z>3$.  We find a large population of objects of
intermediate radio-loudness; there is no evidence in our sample for a
bimodal distribution of radio characteristics.  The sample includes
$\sim 29$ broad absorption line quasars, both high and low ionization,
and a number of new objects with remarkable optical spectra.

\end{abstract}


\section{Introduction}
\label{sectionintro}

The VLA FIRST Bright Quasar survey (FBQS) aims to define a sample of
quasars that bridges the gap between traditional radio-loud and
radio-quiet objects. Given the limiting radio flux density of 1~mJy at
1400~MHz (Becker, White, \& Helfand 1995; hereafter \cite{becker95}),
many of the optically bright quasars discovered by the FIRST survey
fall near the traditional division between radio-loud and radio-quiet
objects.  Previous radio-selected quasar surveys have not reached deep
enough to probe this regime. In a pilot study for the FBQS, Gregg et
al.\ (1996; hereafter \cite{gregg96}) developed criteria to create a
candidate list based on matching the FIRST survey to the Cambridge
Automated Plate Measuring Machine (APM) catalog of POSS-I objects
(\cite{mcmahon92}).  Applying these criteria to the catalog of FIRST
sources from the initial 2682 square degrees surveyed in the north
Galactic cap (White, Becker, Helfand \& Gregg 1997; hereafter
\cite{white97}), a candidate list of 1238 objects with
extinction-corrected, recalibrated magnitudes brighter than 17.8 mag on
the POSS-I $E$ plates has been assembled.  We have now collected
optical spectra for more than 90\% of the candidates and have
identified 467 new quasars in addition to the 169 that were previously
known in this region.

Quasars were originally discovered through their radio emission, but
only $\sim10$\% of them are radio-loud\footnote{We define radio-loud
quasars as those with a radio-loudness parameter $R^*$ greater than 10
(\cite{stocke92}; see \S\ref{sectionradioprop} for further
discussion.)}.  There are now several large surveys for optically
selected quasars that are under way or complete: e.g., the Palomar-Green
survey (\cite{green86}), the Large Bright
Quasar Survey (Hewett, Foltz \& Chaffee 1995; hereafter \cite{LBQS}),
the Edinburgh Quasar Survey (\cite{goldschmidt92}), and the Hamburg/ESO
survey (\cite{wisotzki96,hagen99}).  The FBQS survey will produce a sample of
radio-selected quasar spectra that is comparable in size and quality to
the largest existing optical surveys.  In fact, the catalog published
in this paper already contains more $z>0.2$ quasars brighter than
$B=18$ than does the LBQS (340 for the FBQS vs.~319 for the LBQS)
because even though radio quasars are rarer, the FBQS covers a
substantially larger sky area than the LBQS (2682~deg$^2$
vs.\ 454~deg$^2$ for the LBQS.)

\placefigure{figskydist}

\preprint{
\begin{figure*}
\epsfxsize=\textwidth \epsfbox{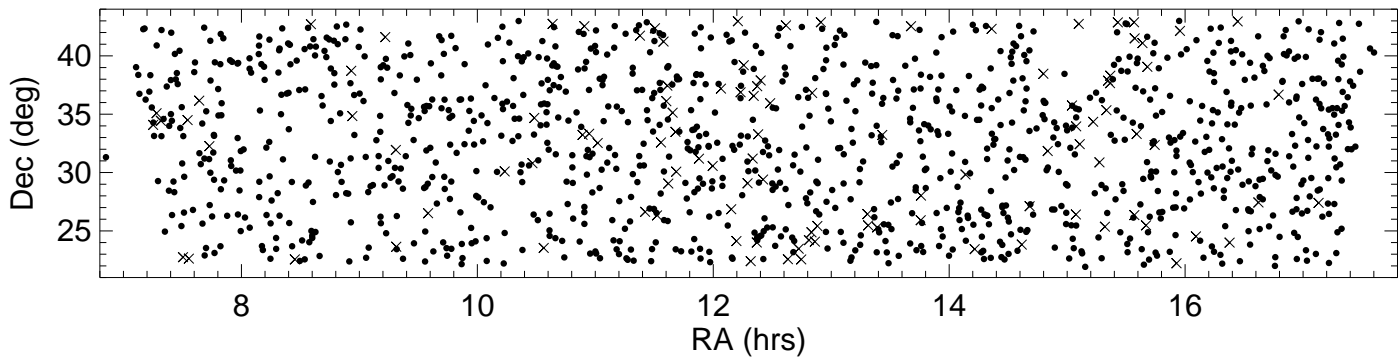}
\caption{
Distribution of the 1238 FBQS candidates on the sky.  Dots are
the 1130 objects identified through spectra; x's mark the 108 objects without
spectra.
}
\label{figskydist}
\end{figure*}
}

In this paper we present this new radio-selected sample of quasars.  We
discuss the criteria used to define the FBQS sample, as well as a new technique
that could be used to make even more efficient samples
(\S\ref{sectionsample}). We describe the optical spectroscopy that was
carried out (\S\ref{sectionspectra}), present the results of the
spectroscopy including spectroscopic classifications for all objects
(\S\ref{sectionresults}), and briefly discuss the results
(\S\ref{sectiondiscussion}).  Detailed analysis of the FBQS sample,
including the spectral properties of the new quasars, will be deferred
to other papers; here our focus is on defining the sample and on
presenting the basic data upon which subsequent work will be based.

\section{The Sample}
\label{sectionsample}

The primary catalog for the FBQS is the VLA FIRST survey which, as of
July 1999, covers $\sim6000$ deg$^2$ (\cite{becker95},
\cite{white97}; see also the FIRST home page at
\url{http://sundog.stsci.edu}).
However, at this time the vast majority of optical
spectroscopy has been restricted to objects drawn from the smaller area
covered by the 1997Apr24 version of the FIRST northern Galactic cap
catalog.  For the purposes of this paper, we will restrict the
discussion to candidates in the north Galactic cap between declinations
of $+22$ and $+43$ degrees, with Right Ascensions ranging from
approximately $\sim7$ to $\sim17$~hours; the area covered is
2682~square degrees.  The spatial distribution of the candidates
included in the paper is shown in Figure~\ref{figskydist}.  The
selection criteria for membership in the sample are:
\begin{itemize}
\item The radio and optical positions must coincide to better than
1.2~arcsec.
\item The recalibrated, extinction-corrected optical magnitude on the
POSS-I red plate $E \le 17.8$.  (The $O$ magnitude roughly corresponds
to $B$ and $E$ to $R$ or to Gunn $r$.)
\item The optical morphology of the object must be stellar on at least
one of the two POSS-I plates.
\item The POSS-I color must be bluer than $O-E=2$.
\end{itemize}
We avoid as far as possible any restriction on the radio properties of
sources in the sample, other than requiring them to be in the FIRST
catalog, which has its own selection effects.  These criteria were
based on the analysis presented in \cite{gregg96} and have been found
to be quite liberal, excluding very few potential quasars.  The
criteria are discussed in detail below.

\subsection{Radio-Optical Positional Coincidence}

We require that the radio and optical sources be separated by no more
than 1.2~arcsec. A discussion of the astrometric accuracy that allows
for such a tight constraint is found in \cite{white97}. The FIRST radio
positions have been used to correct the APM positions for POSS-I plate
distortions before comparing the radio and optical positions
(\cite{mcmahon99}). In general, the required close agreement in
position excludes quasars that do not have at least a weak core radio
component, so the candidate list is biased against
completely lobe-dominated radio sources. The distribution of radio-optical
separations for confirmed quasars is strongly peaked and falls rapidly
beyond 0.5 arcsec (see Fig.~\ref{figsephist}a).  Only 5 (0.8\%) of our
636 quasars have separations between 1.1 and 1.2 arcsec, even though
that annulus contains 16\% of the area searched.  Nonetheless, this
criterion certainly excludes some quasars that are detected by the
FIRST survey.  Figure~\ref{figsephist}(b) shows the fraction of optical
candidates that are found to be quasars as a function of separation; it
declines steadily from $\sim80$\% near 0~arcsec to $\sim20$\% at
1.2~arcsec.  If we make the conservative assumption that the quasar
fraction does not decline further, we estimate that there are $< 10$
additional quasars with separations between 1.2~and 1.5~arcsec that are
excluded from the sample by the separation criterion.

\placefigure{figsephist}

\preprint{
\begin{figure*}
\begin{tabular}{cc}
\epsfxsize=0.45\textwidth \epsfbox{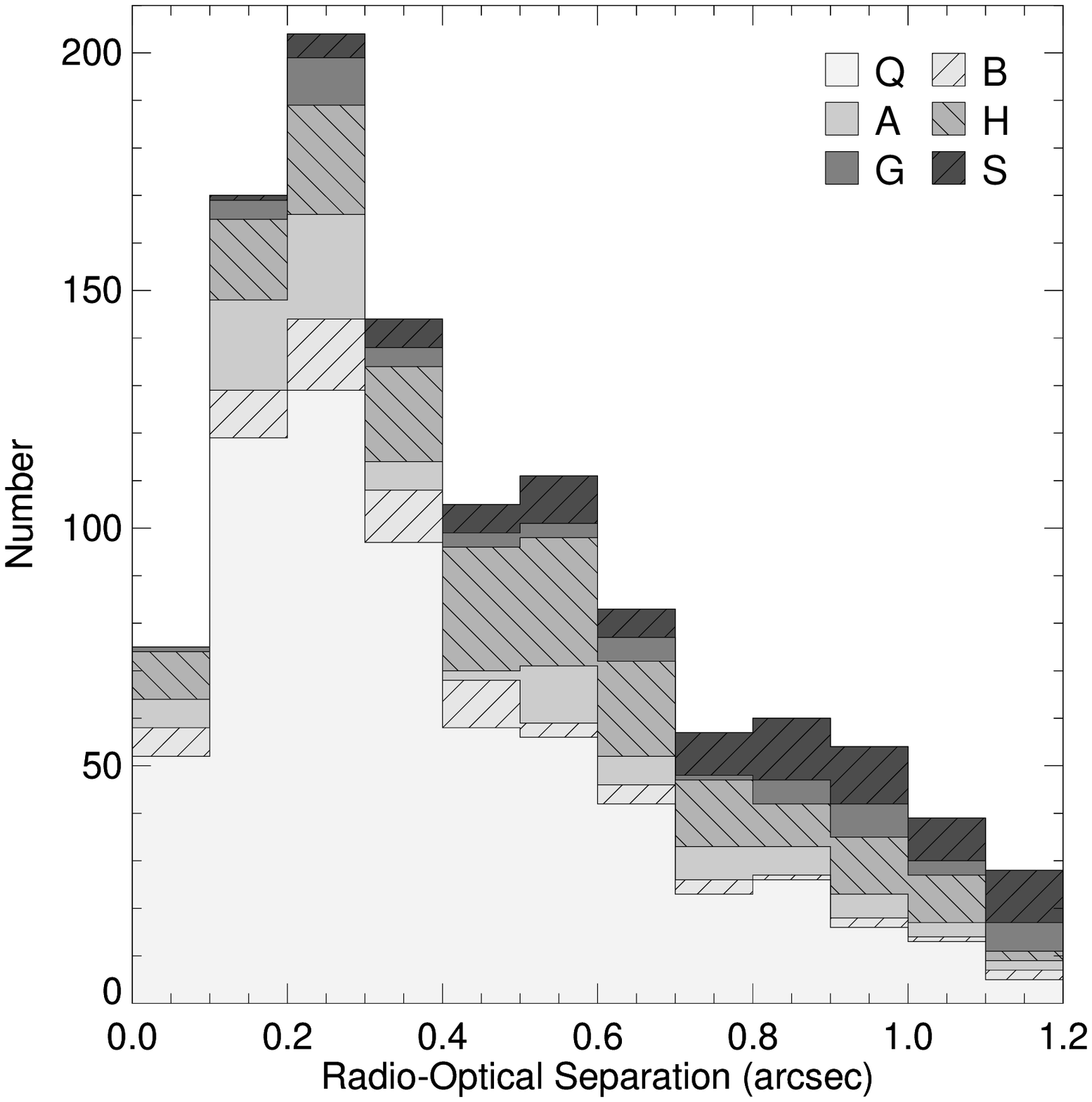} &
\epsfxsize=0.45\textwidth \epsfbox{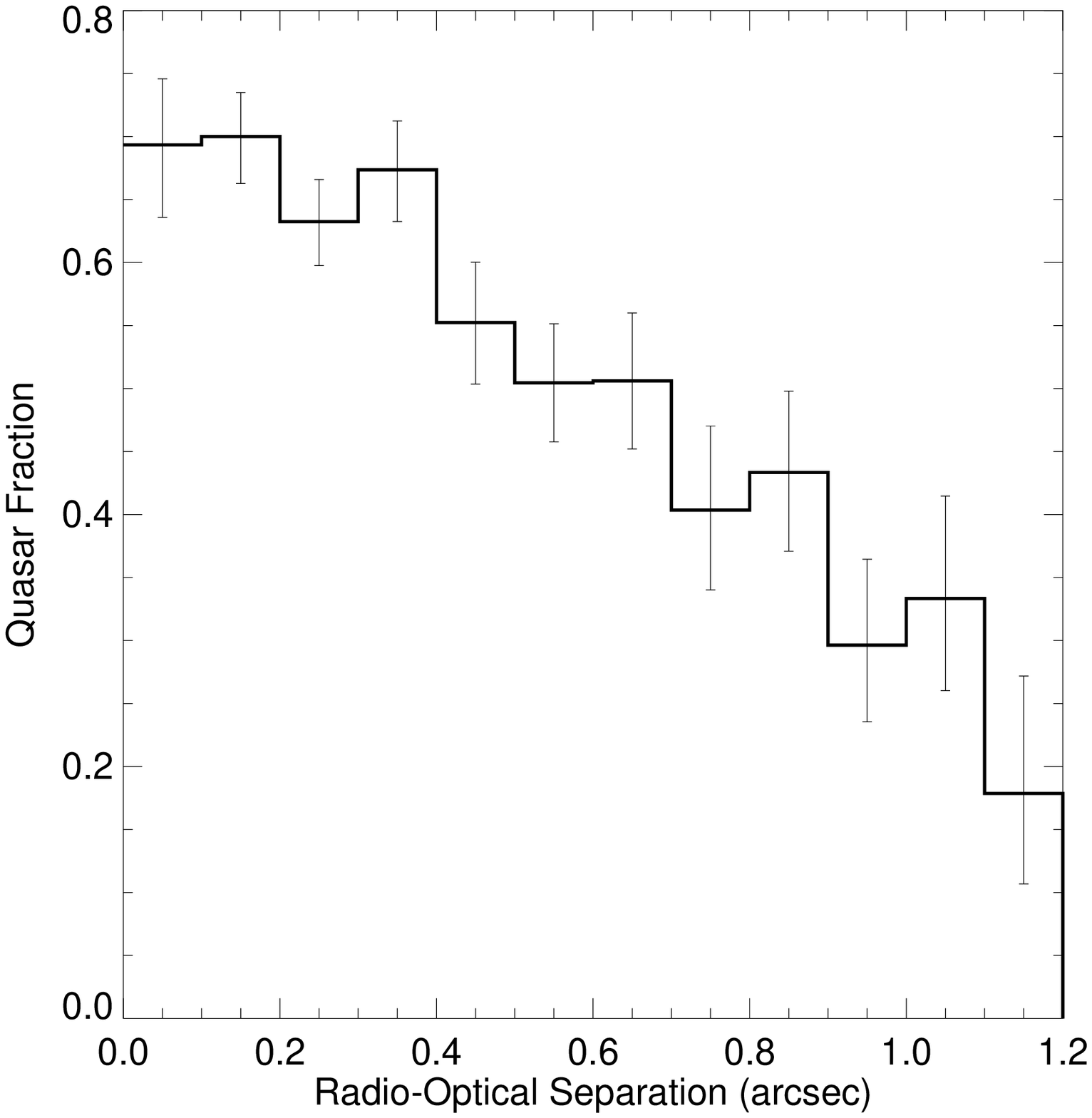} \\
(a) & (b) \\
\end{tabular}
\caption{
(a) Histogram of separations for FBQS candidates identified as quasars
(Q), BL~Lacs (B), narrow-line AGN (A), H~II/star-forming
galaxies (H), galaxies without emission lines (G), and stars (S).  (b)
Fraction of quasar identifications as a function of separation.  Quasars
are strongly concentrated toward small separations.
}
\label{figsephist}
\end{figure*}
}

\subsection{E Magnitude}
\label{sectionmagnitude}

The original candidate list was limited to optical counterparts that
are brighter than 17.5 mag on the POSS-I $E$ plate as measured by the
APM scans. In fact, the APM $E$ magnitude typically overestimates the
actual brightness by $\sim 0.3\pm0.3$ mag, so the original candidate
list went slightly deeper than intended and was not uniform over the
survey area. In an attempt to improve the photometric accuracy and
uniformity of the sample, the APM magnitudes were subsequently
recalibrated plate-by-plate (\cite{mcmahon99}) using magnitudes from
the Minnesota Automated Plate Scanner POSS-I catalog (APS\footnote{The
APS databases are supported by the National Science Foundation, the
National Aeronautics and Space Administration, and the University of
Minnesota, and are available at \url{http://aps.umn.edu/}.},
\cite{pennington93}), which are more uniform than the APM magnitudes
because they were calibrated on a plate-by-plate basis, and a new limit
of 17.8 in $E$ was used to redefine the complete sample.
We preferred not simply to substitute the APS catalog for the APM because
the APS sky coverage is incomplete in the FIRST region and
because the APS catalog excludes objects detected on only one of the
POSS-I plates, which dramatically reduces the fraction of FIRST sources
that have optical identifications since most radio source counterparts
are near the plate limit and so appear on only one plate.  The latter
consideration is not important for the FBQS, but is a disadvantage
for most other optical projects.

After the APS recalibration, we find that the APM magnitudes are
accurate to better than 0.2 magnitudes rms in both O and E
(\cite{mcmahon99}).  This accuracy was determined by comparing the
magnitudes of objects appearing in the narrow overlap regions at the
edges of the POSS-I plates; since the photometric errors are probably
worst at the plate edges, for most objects the errors may be even
smaller.

As a result of the E magnitude adjustments, the sample was augmented by
approximately 100 candidate quasars.  Furthermore, there are instances
of sources in the APS catalog that are not detected on one or both
plates in the APM catalog (the opposite also occurs.)  The most common
reason for such missing sources is that close objects are blended into
a single merged entry.  We added 26 such APS-supplement objects to our
candidate list; these objects are noted in the comments.  Finally, to
improve the uniformity of the sample with Galactic latitude, an
extinction correction was computed for each candidate object using the
$E(B-V)$ map of \cite{schlegel98} with $A(E)=2.7\,E(B-V)$ and
$A(O)=4.4\,E(B-V)$ (estimated from the $E$ and $O$ bandpasses given in
\cite{minkowski63}.) These corrections are usually quite small --- the
median values are $A(E)=0.058$ and $A(O)=0.094$ --- but can be
significant at the high and low RA edges of the survey (see
Fig.~\ref{figextinct}).  We keep all objects that satisfy $E \le 17.8$
and the $O-E$ color cut described below using the extinction-corrected
magnitudes; this adds an additional $\sim90$ candidates to the sample.

\placefigure{figextinct}

\preprint{
\begin{figure*}
\plotone{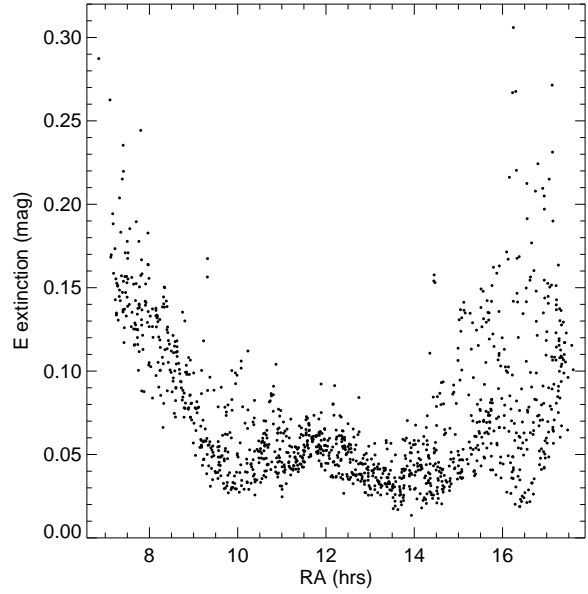}
\caption{
Extinction at $E$ as a function of Right Ascension for FBQS candidates.
The $O$ extinction is 1.63 times larger.  The extinction corrected
magnitudes have been used for the color and magnitude cuts in defining
the FBQS sample.
}
\label{figextinct}
\end{figure*}
}

The extinction-corrected magnitude distribution of identified
candidates is shown in Figure~\ref{figmaghist}(a) for various
classifications.  (The definitions adopted for these object
classifications are given in \S\ref{sectionresults}; briefly, a
``quasar'' is any object with broad emission lines.) The efficiency of
finding quasars is a strong function of magnitude.
Figure~\ref{figmaghist}(b) shows the fraction of candidate objects that
are quasars as a function of $E$ magnitude.  Less than 5\% of the
objects brighter than $E=14$~mag are quasars (most such objects are
brighter than $E=13$ and so do not appear in Fig.~\ref{figmaghist});
this grows to 70\% for candidates fainter than $E=17$~mag.  Most of the
brightest nonquasars are galaxies that the APM has misclassified as
stellar.

\placefigure{figmaghist}

\preprint{
\begin{figure*}
\begin{tabular}{cc}
\epsfxsize=0.45\textwidth \epsfbox{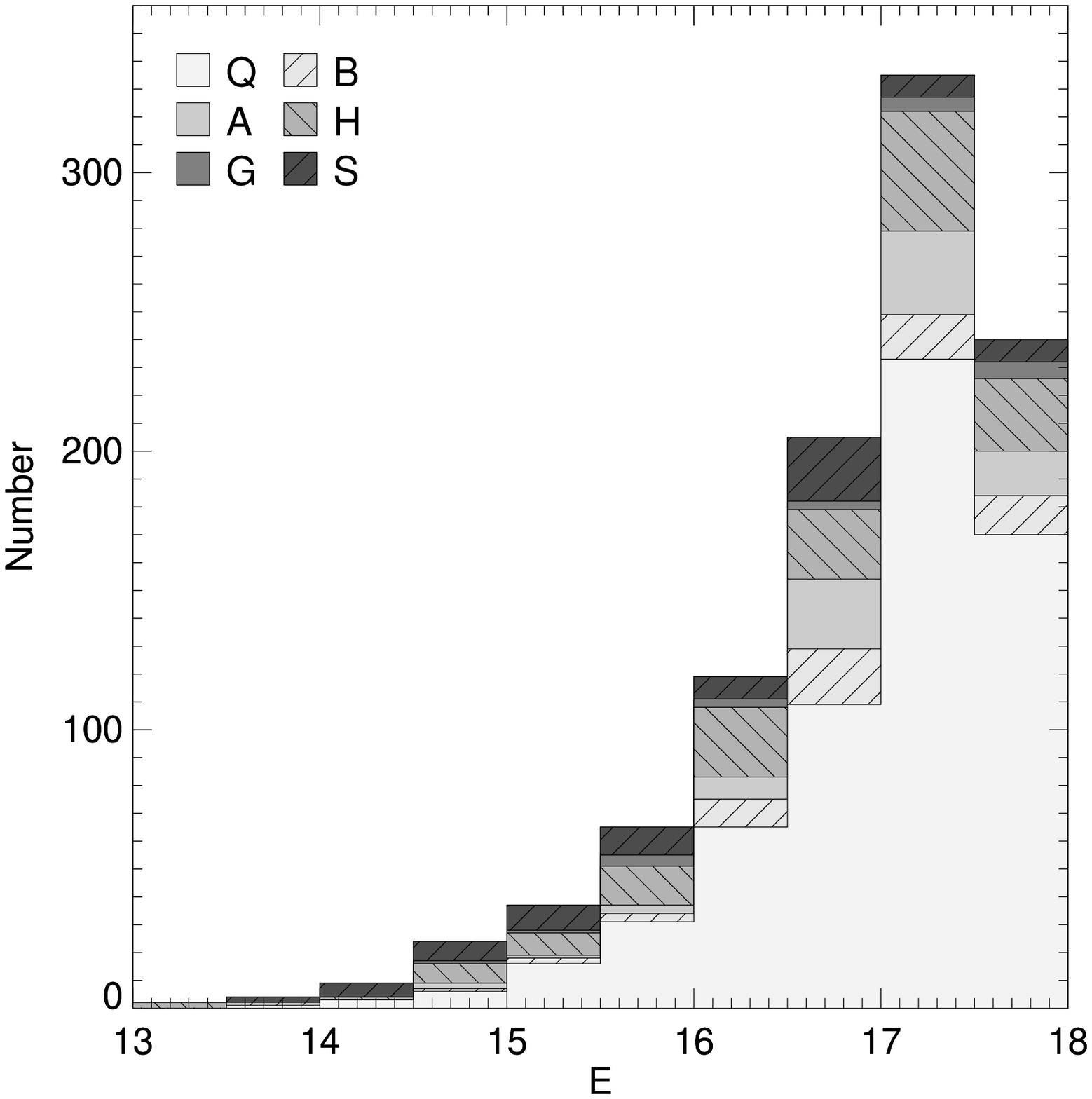} &
\epsfxsize=0.45\textwidth \epsfbox{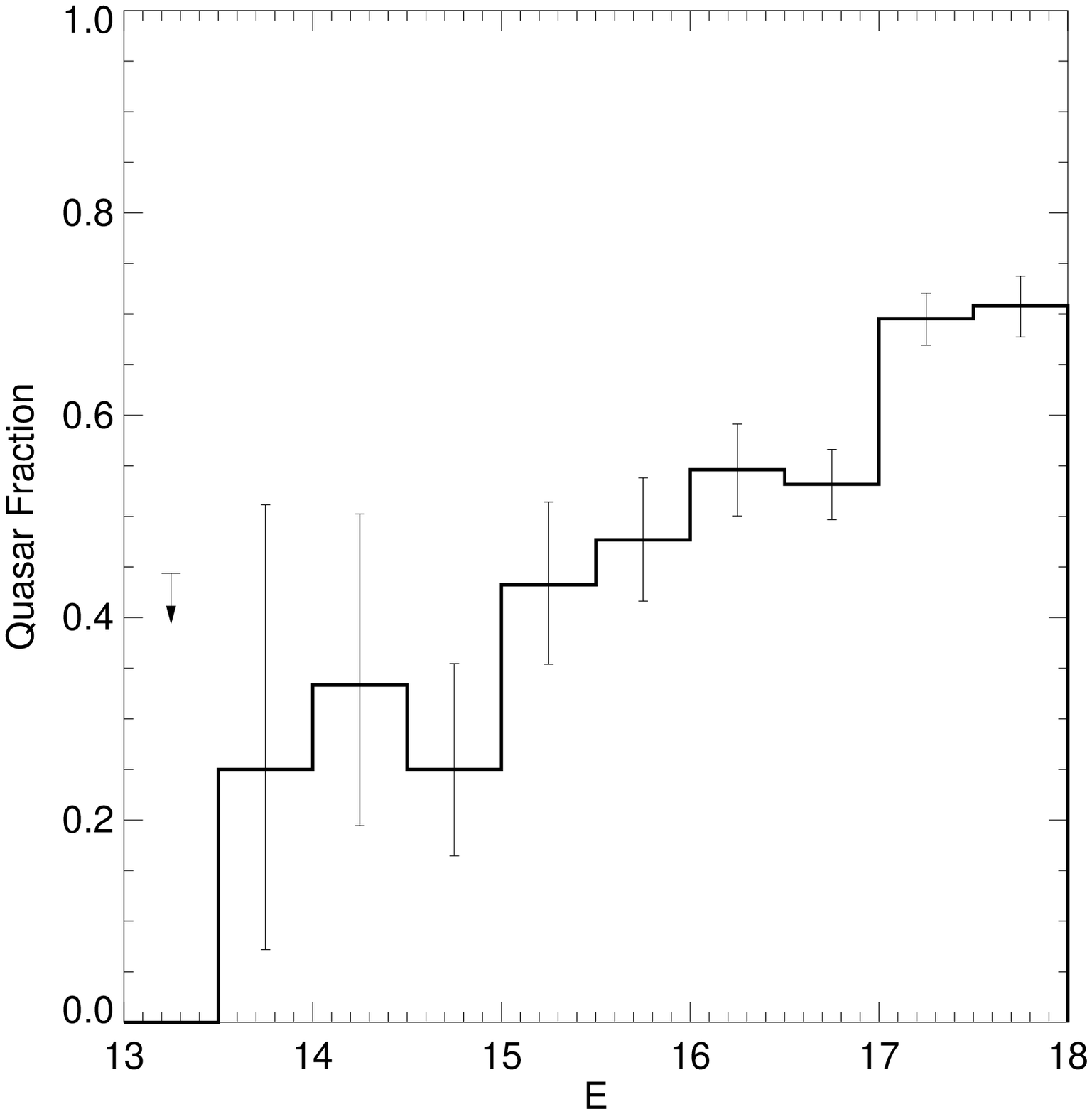} \\
(a) & (b) \\
\end{tabular}
\caption{
(a) Histogram of extinction-corrected $E$ (POSS-I red) magnitudes for
FBQS candidates with identifications.  Note that the 17.5--18.0 magnitude
bin only includes objects between $17.5 < E < 17.8$.
(b) Fraction of quasar identifications as function
of $E$ magnitude.
}
\label{figmaghist}
\end{figure*}
}

The very steep increase in the number of quasars with increasing
magnitude introduces a strong Malmquist bias into this sample (and
other samples of bright quasars.)  There is a large pool of quasars
just below the magnitude threshold, so the sample is certain to include
variable objects that happened to be unusually bright when the POSS-I
plates were taken; it will also include quasars with measurement and/or
calibration errors that made them look brighter than they really are.
When analyzing properties of the sample, readers are cautioned to
consider this bias.  See our study of the variability of a subset of
the FBQS quasars for further discussion of this issue
(\cite{helfandvar}).

\subsection{Optical Morphology}

Since we are searching for quasars, we require all candidates to be
classified by the APM as stellar on at least one POSS-I plate. The APM
star/galaxy separator is not infallible, however. Approximately 14\% of
the quasars found in the FBQS are classified as nonstellar on one of
the two POSS-I plates.  Since we admit Seyfert galaxies into our quasar
sample, some of these nonstellar classifications may be correct.  Fully
75\% of the objects classified as stellar on both plates turned out to
be quasars. In comparison, only 25\% of the objects classified as
stellar on only one plate proved to be quasars.

Nonetheless, it is likely that some quasars are being lost to the
sample because they are being misclassified as galaxies on both
plates.  A match between the APM catalog and the \cite{veron98} quasar
catalog found that 10\% of the V\'eron quasars brighter than $\sim17.8$
in $E$ were classified as galaxies on both POSS-I plates by the APM
(\cite{gregg96}).  However, $\sim40$\% of such objects are identified
as Seyfert galaxies in the V\'eron catalog, so that the APM galaxy
classification is likely to be correct.  The FBQS sample is certainly
not complete for Seyfert galaxies due to the restriction to point-like
optical sources.  This morphology criterion similarly makes the FBQS
sample somewhat incomplete at low redshifts ($z \lesssim 0.2$), where
the quasar host galaxy may be bright enough to skew the APM
classification.

Requiring a point-like optical morphology may also discriminate against
another interesting class of quasar: gravitationally lensed (or binary)
objects.  Gravitational lenses with two roughly equally bright images
separated by $\sim2$~to $\sim10$~arcseconds are likely to be recorded
as a single non-stellar object in the APM catalog; they are also likely
to have less accurate positions in both the radio and the optical due
to their extent.  Such wide lenses are probably best identified using
higher-resolution images than can be obtained from photographic Schmidt
surveys.  However, most gravitationally lensed quasars have smaller
separations than this and so will be included in our sample despite not
being true point sources.  In an imaging search for lenses among the
FBQS sample (\cite{schechter98}), we have so far found two lensed
objects; both have component separations of $\sim1$~arcsec or less and
are unresolved by the APM, which classifies each as stellar on both
plates.  We have also identified one binary quasar (\cite{brotherton99})
with a separation of 2.3~arcsec, which is similarly classified as
stellar on both plates, although it is slightly too faint to be in the
FBQS sample.

\subsection{Optical Color}
\label{sectioncolorcut}

Lastly, we excluded FIRST sources with very red optical counterparts
($O-E > 2$) since in the pilot study no quasars were this red.  At the
risk of missing a few intrinsically red or high-redshift quasars, this
cut substantially reduces the number of candidates to observe. A
histogram of the colors of all the quasars identified to date
(Fig.~\ref{figcolorhist}a) shows very few close to the color limit and
a very low discovery efficiency near $O-E = 2$
(Fig.~\ref{figcolorhist}b). There are only 7 confirmed quasars with
dereddened $O-E \ge 1.8$.  Although we cannot exclude a large
population of red quasars, there are very few quasars near our color
boundary, whereas the non-quasar contamination of the candidate list
has a strong dependence on color. Figure~\ref{figcolorhist}(a) also
indicates the color distribution of the BL Lacs, AGN, H~II/star-forming
galaxies, and normal galaxies identified in our survey.

\placefigure{figcolorhist}

\preprint{
\begin{figure*}
\begin{tabular}{cc}
\epsfxsize=0.45\textwidth \epsfbox{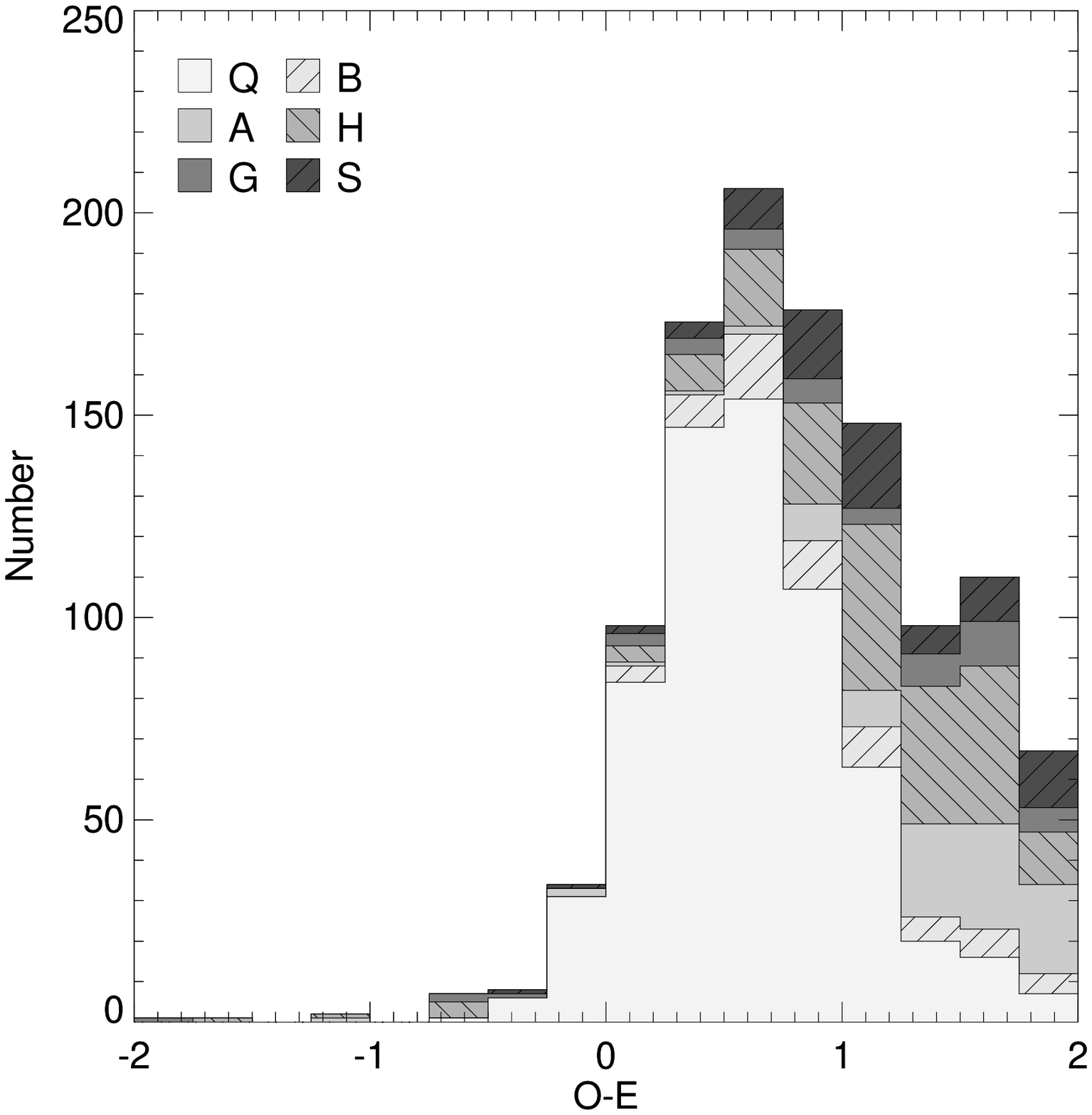} &
\epsfxsize=0.45\textwidth \epsfbox{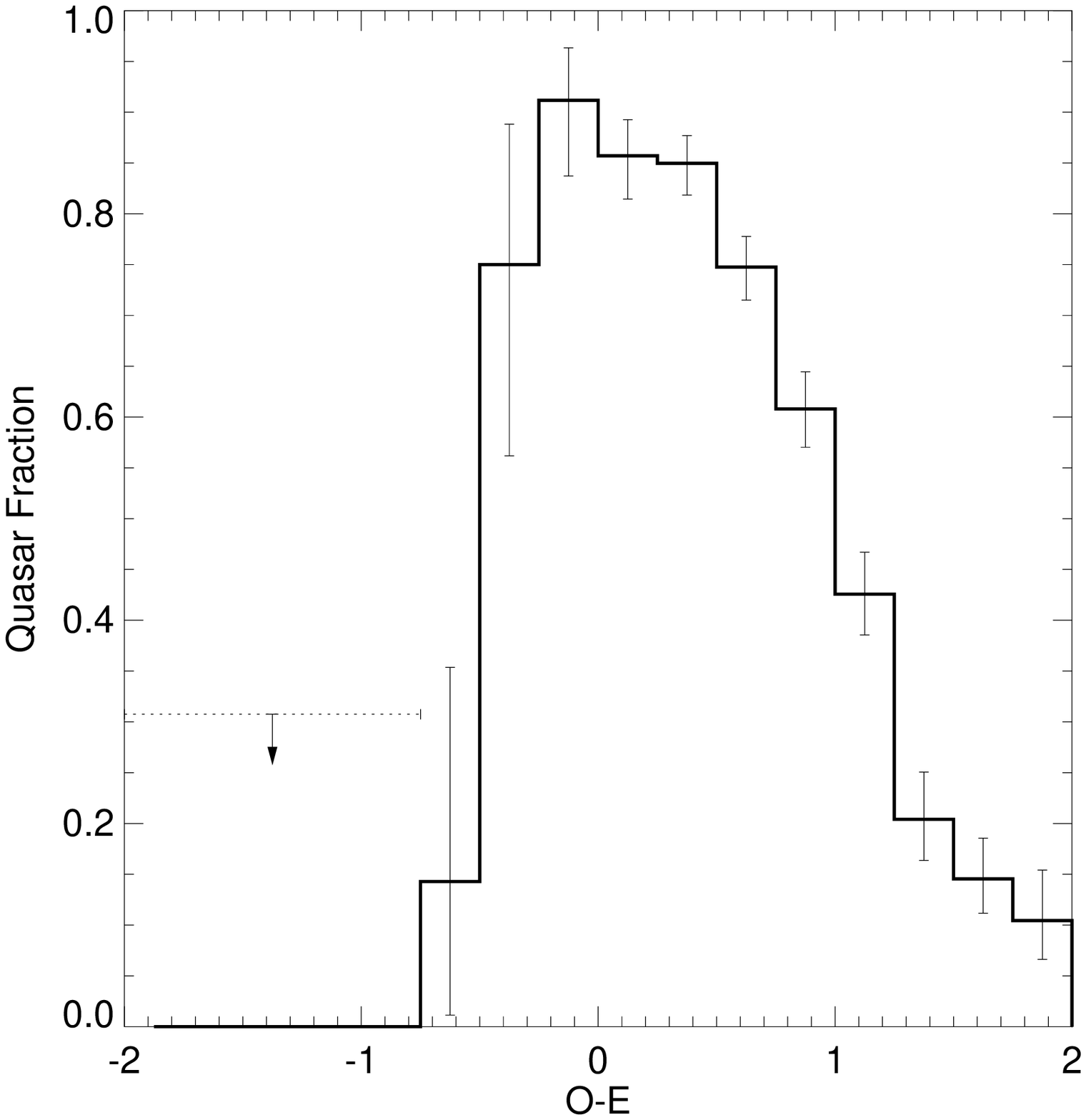} \\
(a) & (b) \\
\end{tabular}
\caption{
(a) Histogram of extinction-corrected colors of identified FBQS
candidates.
(b) Fraction of quasar identifications as function
of color.
}
\label{figcolorhist}
\end{figure*}
}

Our color cut does discriminate against one known class of red
quasars:  very high redshift quasars ($z \gtrsim 3.5$).  \cite{hook95}
describe the evolution of APM/POSS-I colors with redshift and show that
when the Lyman forest absorption moves into the $O$ band, the $O-E$
color becomes dramatically redder.  This trend is visible in a plot of
$O-E$ versus $z$ for the FBQS quasars (Fig.~\ref{figcolorz}), where the
colors of $z>3$ quasars are distinctly redder than those at lower
redshifts.  A prime goal of extending the FBQS sample to include redder
objects would be to find high redshift quasars (e.g., \cite{hook98}),
but in view of the contamination of the red objects in the APM catalog
by galaxies, due mainly to the poor morphology discrimination feasible
with the POSS-I plates, we decided it was best for the FBQS to postpone
a red quasar search until we have higher quality optical images (e.g.,
from the Sloan Digital Sky Survey and other deeper, higher resolution
CCD surveys.)

\placefigure{figcolorz}

\preprint{
\begin{figure*}
\plotone{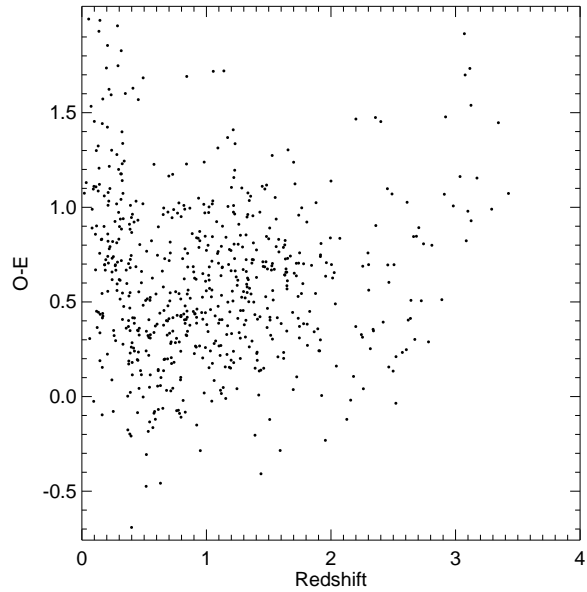}
\caption{
Optical color versus redshift
for FBQS quasars.
At $z>3$ the colors of the quasars begin to redden, so that
high-$z$ quasars may be lost from the FBQS sample due to the
color cut at $O-E \le 2$ (\cite{hook95}).
}
\label{figcolorz}
\end{figure*}
}

In any case, few if any high redshift quasars were missed as a
result of our color cut.  The limit at $O-E=2$ is substantially redder
than the typical unreddened quasar color ($O-E \sim 0.5$,
Fig.~\ref{figcolorhist}), so it does allow for significant absorption
in the $O$ band; and we do not expect to find many $z>3.5$ quasars
brighter than $E=17.8$~mag simply because of the large ultraviolet
luminosities required.  If there is a closer population of quasars that
is heavily reddened by either intrinsic or intervening dust absorption,
the absorption at $E$ would have to be several magnitudes in order to
produce a differential extinction of a magnitude or more in $O-E$.
Consequently, such objects would also be too faint to be included in
this bright sample.

\subsection{Radio Properties}
\label{sectionradioprop}

There are no cuts to the sample based on radio flux density or radio
morphology (other than the unavoidable requirement that the radio
source have at least a weak core component, as mentioned above.)
However, the FIRST catalog itself introduces several minor selection
effects (\cite{becker95}; \cite{white97}).  The FIRST sensitivity limit is
somewhat non-uniform on the sky, with small ($\sim15$\% peak-to-peak) variations due
to the observing strategy and large variations due to decreased
sensitivity in the vicinity of bright sources.  The fraction of the
survey area affected by sensitivity variations is small:  86\% of the
FBQS area has a $5\sigma$ flux density limit of $\le 1$~mJy (the
catalog has a hard limit at 1~mJy even when the images are slightly
better than that), and 98\% of the survey has a limit $<1.25$~mJy.  The
FIRST coverage map (\cite{white97}; available through the FIRST web
pages) gives the sensitivity as a function of position.

The FIRST survey detection limit applies to the peak flux density of
sources rather than to the integrated flux density.  Consequently,
extended sources with total fluxes greater than 1~mJy may not appear in
the catalog because their peaks fall below the detection threshold.
For the objects selected in the FBQS, we expect the insensitivity of
FIRST to extended sources to contribute less to the incompleteness than
the requirement that FBQS quasars must have a nuclear radio component
in order to meet the positional coincidence criterion.  The latter, for
example, makes the FBQS incomplete to the population of
radio quasars that have extended, fossil radio lobes but no active
nuclear emission (which must exist, though they are rare.)

\placefigure{figfluxhist}

\preprint{
\begin{figure*}
\begin{tabular}{cc}
\epsfxsize=0.45\textwidth \epsfbox{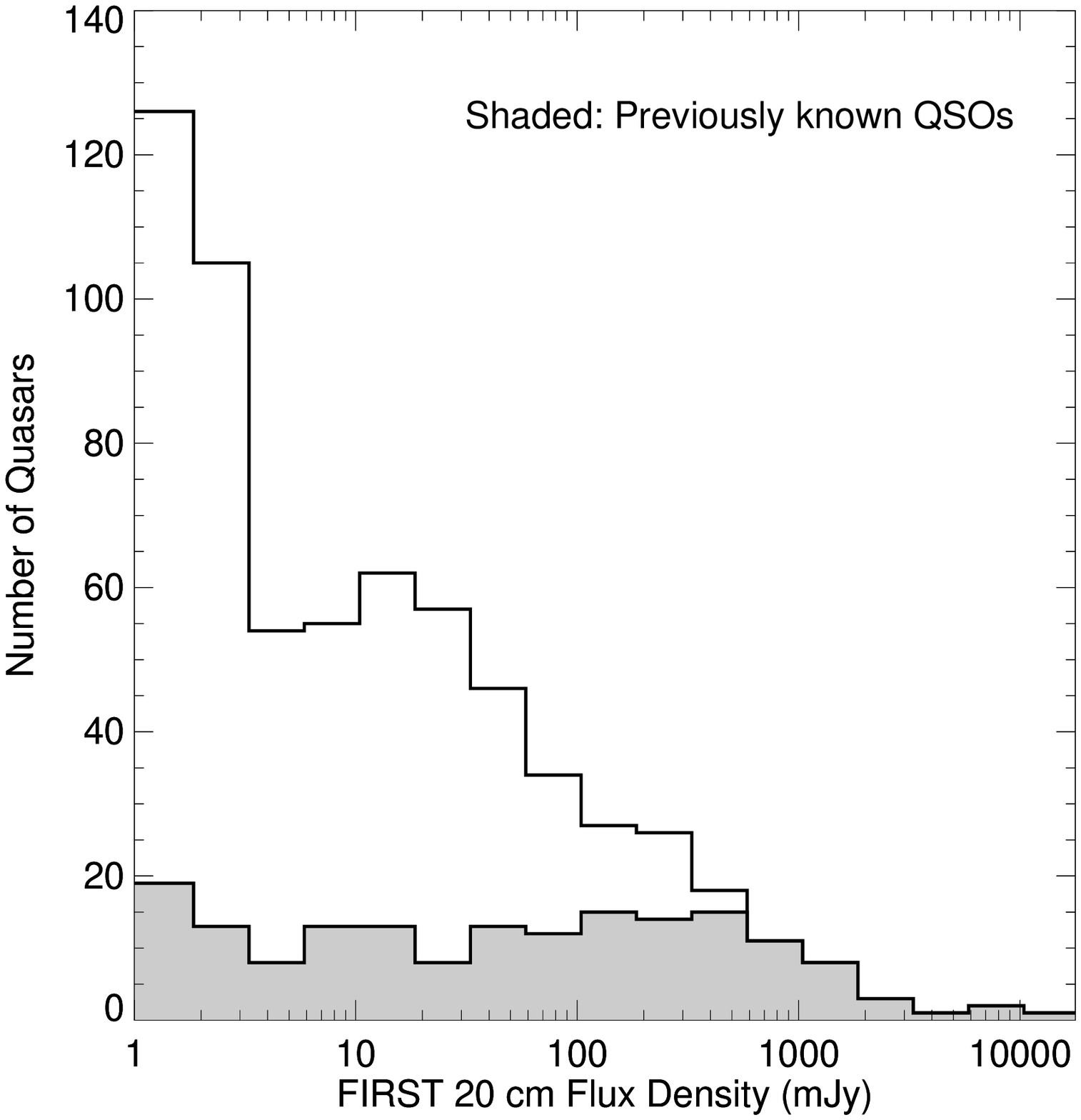} &
\epsfxsize=0.45\textwidth \epsfbox{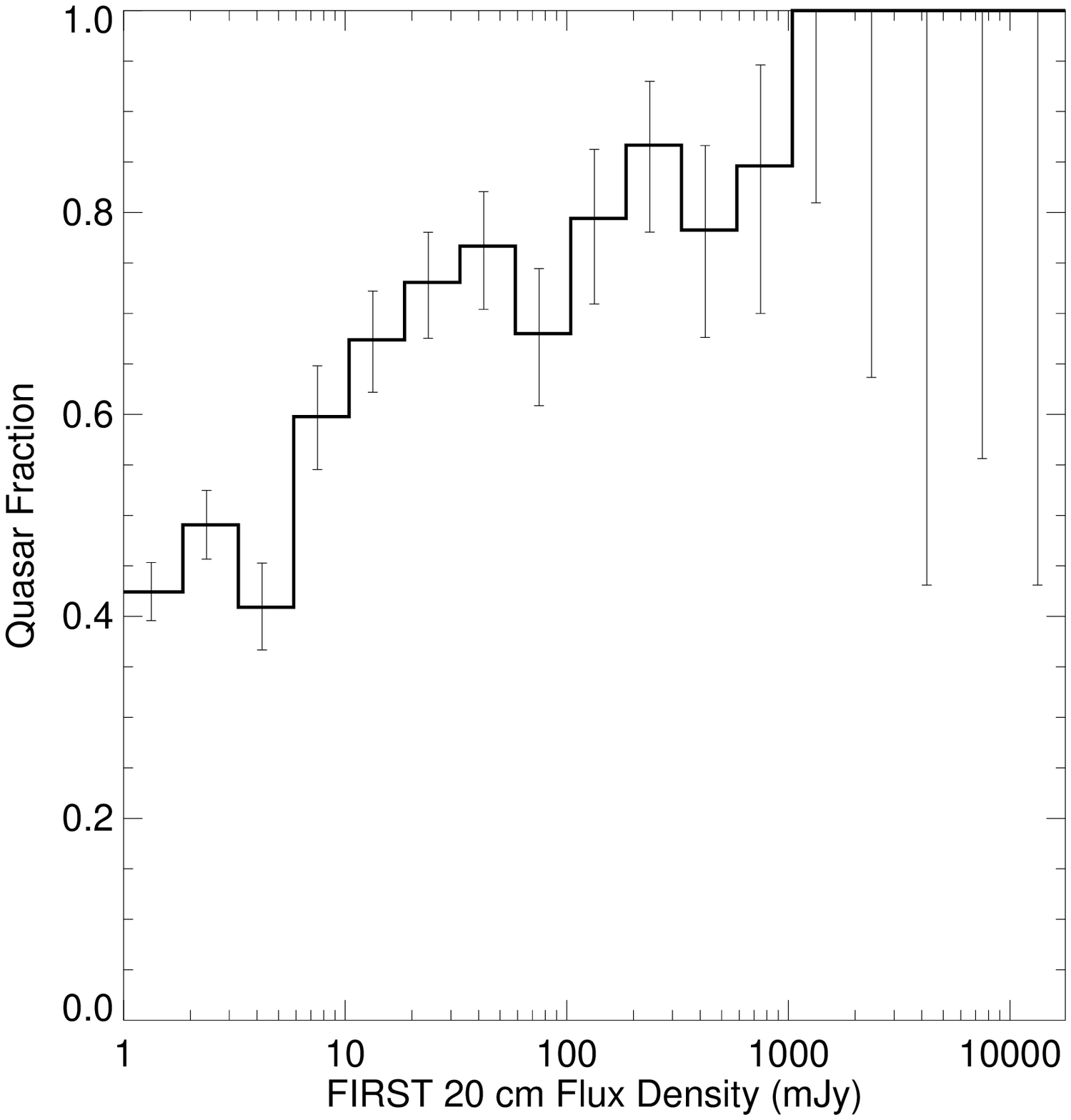} \\
(a) & (b) \\
\end{tabular}
\caption{
(a) Histogram of FIRST integrated radio flux densities
of FBQS quasars.  Shaded: previously known quasars.
(b) Fraction of FBQS candidates identified as quasars as function
of radio flux density. The flux densities come from the FIRST
catalog and so include only the core radio emission.
}
\label{figfluxhist}
\end{figure*}
}

\placefigure{figlumz}

\preprint{
\begin{figure*}
\begin{tabular}{cc}
\epsfxsize=0.45\textwidth \epsfbox{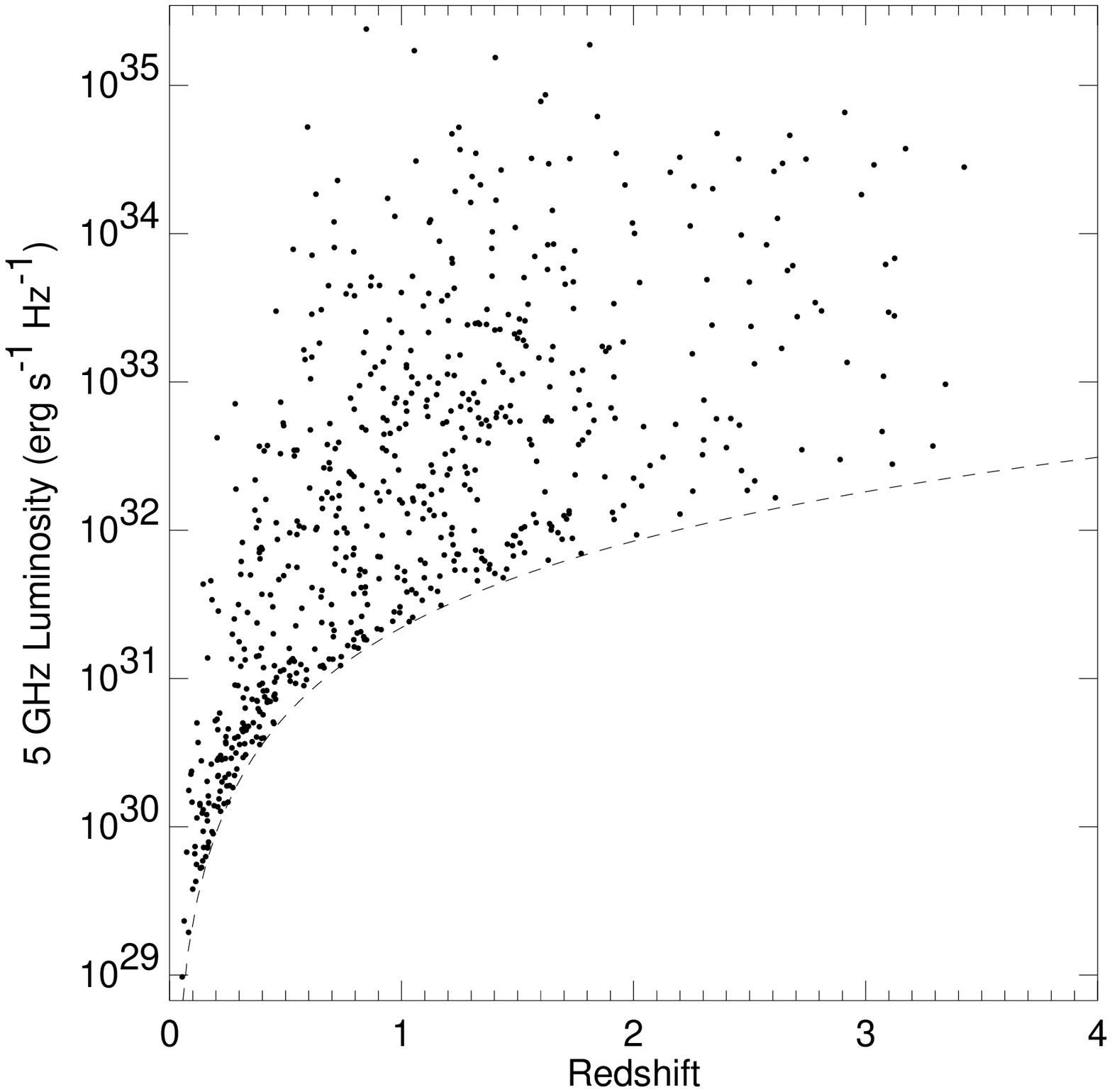} &
\epsfxsize=0.45\textwidth \epsfbox{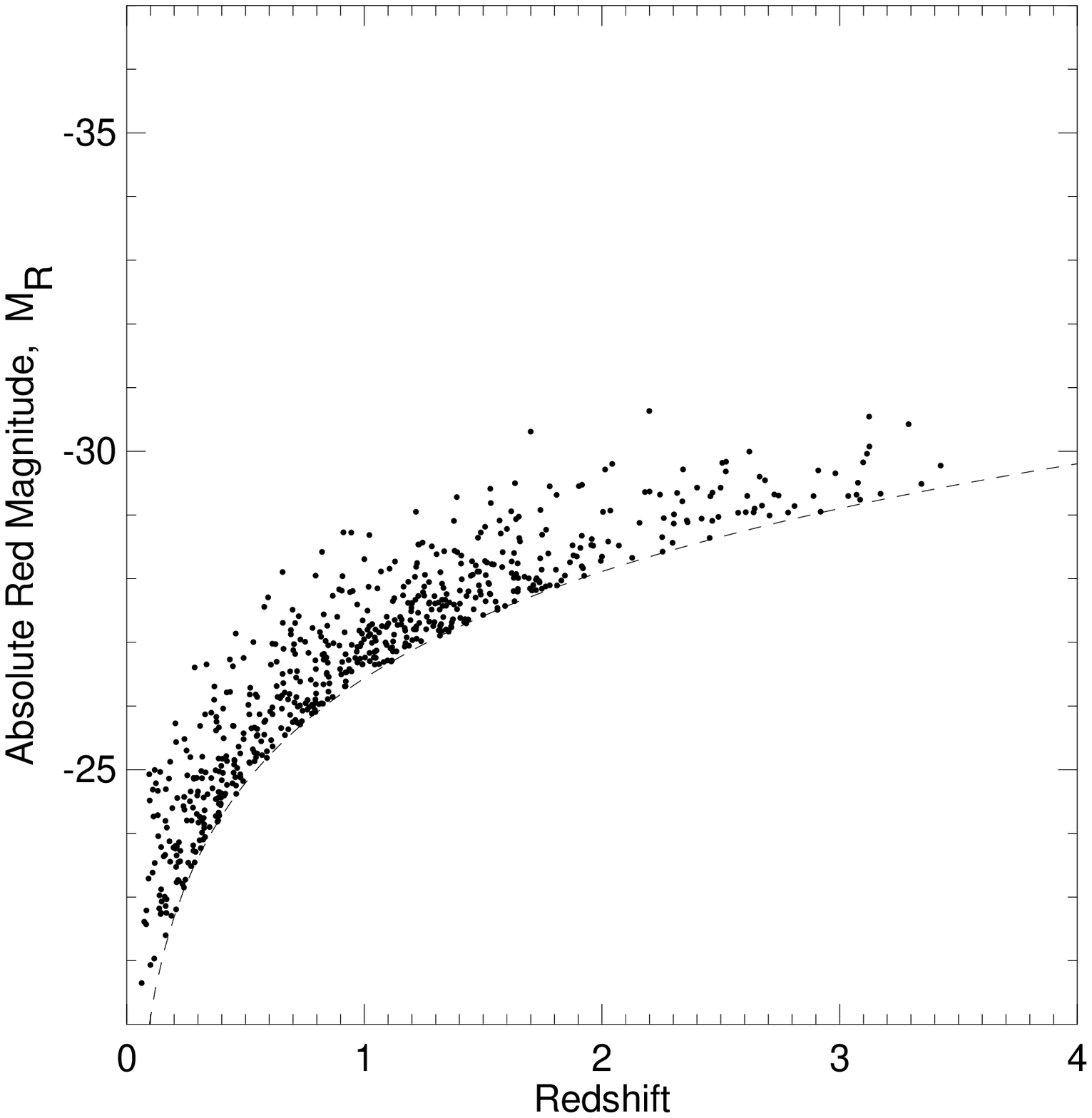} \\
(a) & (b) \\
\end{tabular}
\caption{
FBQS quasar luminosities versus redshift in the radio and optical.
(a) Radio luminosity at a rest frequency of 5~GHz versus redshift,
using spectral index $\alpha=-0.5$.
(b) Absolute red magnitude versus redshift, using spectral index
$\alpha_{opt}=-1$ for the $K$-correction.  Dashed lines show 1~mJy
FIRST detection limit and $E=17.8$ APM magnitude limit.  The radio
luminosities have a much larger dynamic range and do not crowd as
closely against the detection limit as do the optical magnitudes.
}
\label{figlumz}
\end{figure*}
}

The FBQS sample constitutes a significantly different population of
quasars than has been seen in previous radio surveys.
Figure~\ref{figfluxhist} shows the distribution of radio flux densities
for the confirmed quasars.  The number of quasars increases
dramatically with decreasing radio flux density near the FIRST survey
limit. This is because at 1~mJy, the FIRST survey is probing the
larger, radio-quiet (but obviously not radio-silent) quasar
population.  This is also clearly seen in Figure~\ref{figlumz}(a),
which plots the monochromatic radio luminosity versus redshift for FBQS
quasars.  FIRST is sensitive to a large population of low
radio-luminosity quasars, especially at low redshift.

In \S\ref{sectionmagnitude} we discussed the importance of the
Malmquist bias vis a vis the E magnitude cutoff for the sample.  The
rapid increase in the FBQS counts toward our 1~mJy flux density cutoff
implies that there is also a radio Malmquist bias in our sample.  The
slight nonuniformity of the FIRST survey flux density limit directly
affects the FBQS sample and could create the appearance of large scale
structure, since regions of higher sensitivity will have an excess of
quasars.  This effect must be accounted for in analyses of large-scale
structure in the FBQS sample.

Even though the number of quasars increases sharp\-ly below a flux
density of several mJy, the number of non-quasar candidates rises even
faster such that the efficiency of identifying quasars decreases as the
radio flux density decreases (see Fig.~\ref{figfluxhist}b).
Nonetheless, quasars near the FIRST flux density limit are worth the
effort because it is here that the FBQS is making a unique
contribution, probing the transition between radio-loud and radio-quiet
objects.  The distribution of objects in the transition region is
discussed further below (\S\ref{sectionbimodal}).

\subsection{How Efficiently Can Quasars Be Selected?}

The fraction of FBQS candidates that turns out to be quasars is, as
shown above, a
function of magnitude, color, radio-optical separation, etc.
This raises an interesting question: how efficiently
can we select quasars, using only the information that is available
{\it before} taking spectra?  In other words, how well can we maximize
the number of quasars discovered per spectrum taken?

It is clear that by setting tight limits on the separations
(Fig.~\ref{figsephist}) and colors (Fig.~\ref{figcolorhist}), and by
observing only the optically fainter (Fig.~\ref{figmaghist}) and
radio-brighter (Fig.~\ref{figfluxhist}) objects, we could eliminate
many of the FBQS candidates that turn out not to be quasars and so
increase our efficiency.  However, the completeness of the sample would
suffer greatly using such a strategy, and many of the quasar candidates
rejected would be among the more interesting types (e.g., radio-quiet,
radio-intermediate, and high-redshift objects.)

To explore this question quantitatively, we have applied
artificial-intelligence methods to classify the FBQS candidates
according to their {\it a priori} (before taking spectra) probability
of being quasars, $P(Q)$.  There are a number of classification methods
that we could have used (neural nets, nearest-neighbor algorithms,
clustering methods, Bayesian classifiers, etc.); we chose to use the
oblique decision tree classifier OC1\footnote{OC1 is available via
anonymous ftp from
\url{http://www.cs.jhu.edu/salzberg/announce-oc1.html}}
(\cite{murthy94}), which we have found to represent a good compromise
between the competing goals of being fast in training and in
application and of generating accurate, understandable classification
algorithms.  In the past this method has been successfully applied to
cosmic ray identification in HST images (\cite{salzberg95}), to
star-galaxy discrimination for the Guide Star Catalog-II
(\cite{white97b}), and to the problem of flagging sidelobes appearing
in the FIRST catalog (\cite{white97}).

\subsubsection{Description of OC1}

OC1 takes as input a {\it training set} of objects that have known
classifications; in our case, we use all objects with spectra as the
training set.  Each object is described by a vector of numerical {\it
features} (magnitudes, colors, and so on.) OC1 constructs a {\it
decision tree} that accurately classifies the training set.

At each node in the tree a linear combination of the feature values
plus a constant is computed; if the sum is positive, the right branch
of the tree is taken, otherwise the left branch is taken.  After each
branch, there may be another decision test node, or the tree may
terminate (a ``leaf'' node).  To classify an unknown object, one starts
at the tree's root node and works down the decision tree, doing a
series of tests and branches, until a leaf node is reached.  The OC1
program attempts to produce trees that have pure samples of
training-set objects at each leaf node (i.e., in the current case it
tries to separate the training objects so that each leaf has nearly all
quasars or nearly all non-quasars.) The unknown object is classified as
quasar or not depending on which is the majority type of object at the
final leaf.

\subsubsection{Voting Decision Trees}

We have improved on the accuracy of the classification by using not
just a single tree, but rather a group of 10 trees that vote.  This
multiple-tree approach has been shown to be quite effective
(\cite{heath96}).  Searching for a good decision tree is an extremely
difficult optimization problem; to find an approximate solution, OC1
uses a complex search algorithm that includes some randomization to
avoid the classic problem of getting stuck in local minima in the
many-dimensional search space.  Thus, one can run OC1 many times using
different seeds for the random number generator to produce many
different trees.

\cite{heath96} used a simple majority voting scheme:  classify the
object with each tree and then count the number of votes for each
class.  We have improved on this by using a weighted voting scheme,
where each tree splits its vote between the quasar and non-quasar
classes depending on the populations of the two types from the training
set at that leaf.  If an object winds up at a leaf node with $N$
training set objects of which $Q$ are quasars, the tree's single vote
is split into $(Q+1)/(N+2)$ in favor and $(N-Q+1)/(N+2)$ against a
quasar classification.  (The particular form used was derived from the
binomial statistics at the leaf.)  We considered a variety of other
voting schemes, but this one appeared to be the most accurate and
robust.

\subsubsection{Testing the Classification Accuracy}

The classifier was tested using 5-fold cross-valida\-tion.  The
training set (which consists of $\sim1000$ objects with spectra) is
divided into 5 equal-sized, randomly selected subsets (or {\it
folds\/}).  Setting aside the first fold, 10 decision trees are
constructed by training on the other 4 folds.  Then the trees are
tested on the first fold, which was not used in the training, computing
$P(Q)$ for each object in the fold.  This process is repeated 5 times,
each time holding back a different fold; when complete, we have used
each object for testing exactly once.  This technique, which is a
standard approach, avoids the over-optimistic results for
classification accuracy one would get if one simply trained on all the
data and then tested the classifier on the same data.

\subsubsection{Results}

The approach may sound complicated, but the ultimate result is simple
--- for each object, we compute the probability $P(Q)$ that the object
is a quasar.  A perfect classifier would assign all quasars $P(Q)=1$
and all non-quasars $P(Q)=0$; in a non-ideal case, we want to see
quasars with $P(Q)$ values concentrated near unity and non-quasars near
zero.

\placefigure{figpqhist}

\preprint{
\begin{figure*}
\begin{tabular}{cc}
\epsfxsize=0.45\textwidth \epsfbox{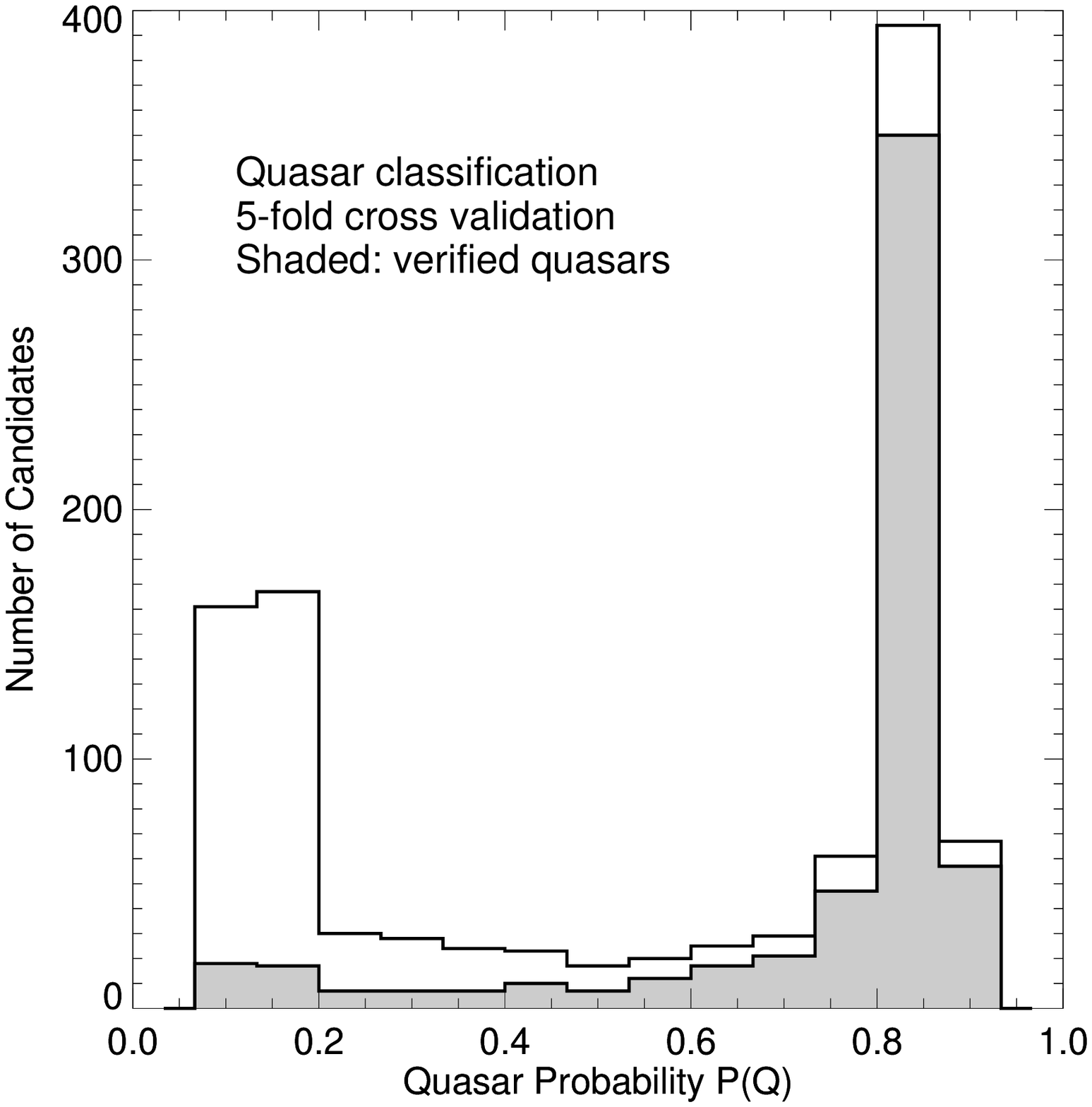} &
\epsfxsize=0.45\textwidth \epsfbox{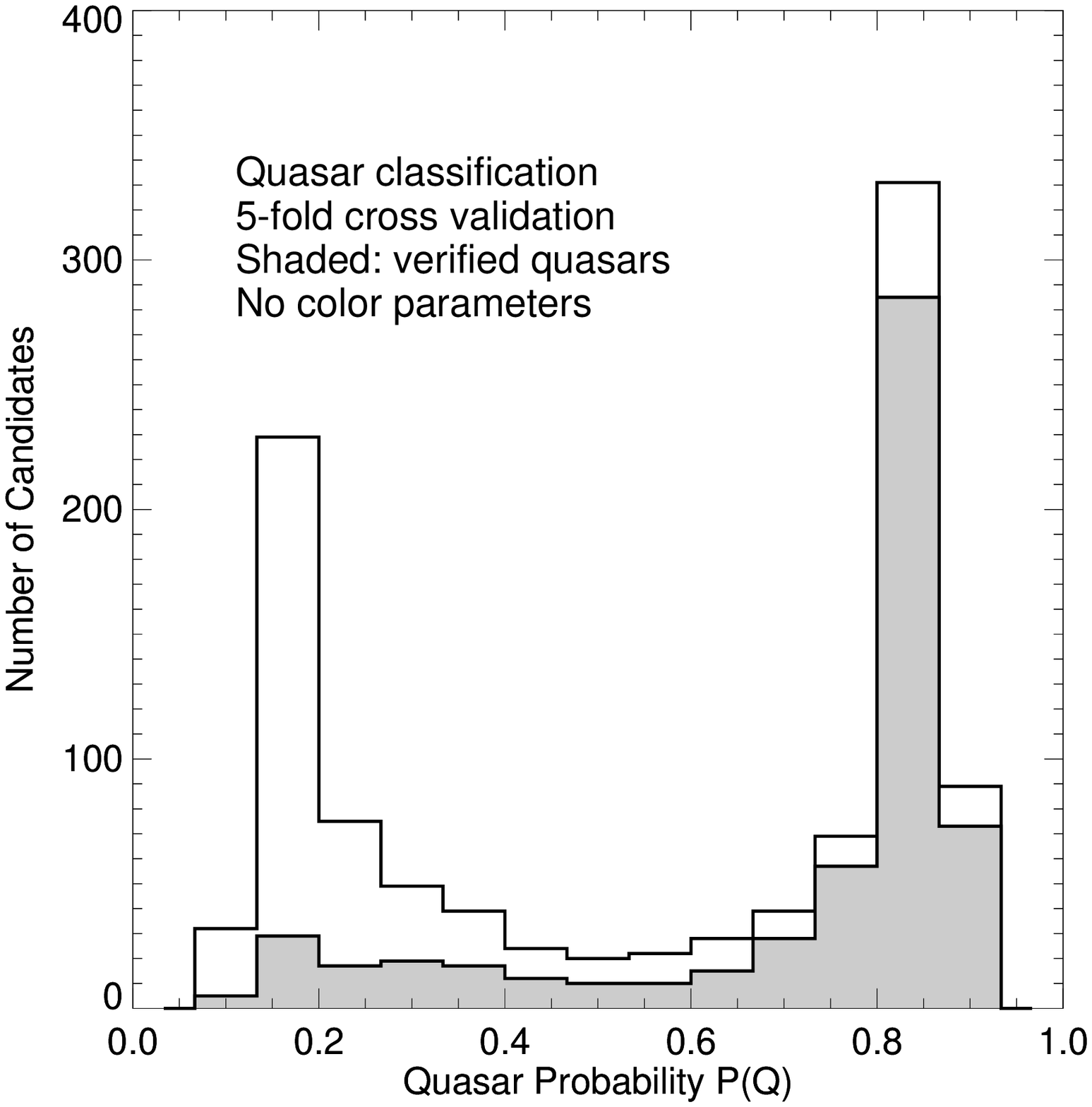} \\
(a) & (b) \\
\end{tabular}
\caption{
Distribution of the {\it a priori} probability that an object is a
quasar, $P(Q)$, for all identified FBQS candidates.  The shaded
histogram is the distribution of confirmed quasars, while the unshaded
area represents non-quasars.  A perfect classifier would put all the
quasars at $P(Q)=1$ and all the non-quasars at $P(Q)=0$.  The
probabilities were calculated with decision trees using 5-fold
cross-validation, as described in the text.  (a) $P(Q)$ computed using
all available parameters, including $O-E$ color.  (b) $P(Q)$ computed
without color.  The second approach is less efficient at distinguishing
quasars from non-quasars, but it does not discriminate against
high-redshift or heavily reddened broad absorption line quasars as does
the method including color.  In both cases, the great majority of the
$P(Q)<0.2$ quasars are at low redshift ($z<0.3$) and are presumably
slightly resolved.
}
\label{figpqhist}
\end{figure*}
}

Figures~\ref{figpqhist}(a) and \ref{figpqhist}(b) show the distribution
of $P(Q)$, the probability that an object is classified as a quasar,
using 10 voting trees.  The distinction between the two cases is the
set of features used for classification.  Example (a) used 7
parameters: $E$, $O-E$, $\log S_p$ (where $S_p$ is the FIRST peak flux
density), $S_i/S_p$ (where $S_i$ is the FIRST integrated flux density),
the radio-optical separation, and the two PSF parameters from the APM
catalog that measure how point-like the object is on the $O$ and $E$
plates.  Example (b) omitted the parameter $O-E$, so that no color
information was included.

Both panels of Figure~\ref{figpqhist} clearly show that the classifier
does a good job of separating the FBQS candidates into high and low
probability quasars.  In both cases the distribution is encouragingly
close to that of an ideal classifier.  The quasar/non-quasar separation
is somewhat better when the $O-E$ color information is used: the
distribution is more strongly peaked at the low and high $P(Q)$ ends.
We created the color-free classifier because an examination of the
quasars with $P(Q) < 0.2$ (the shaded bins below 0.2 in
Fig.~\ref{figpqhist}a) shows that 4 of the 36 misclassified objects are
in the interesting category of red quasars, either because they are
high redshift (J092104.4+302031, $z=3.34$)
or because they have strong
broad absorption lines (BALs; J105427.1+253600,
J120051.5+350831, and J132422.5+245222).
If we exclude color as a classification criterion,
the number of $P(Q)<0.2$ quasars stays about the same (34 objects), but
the misclassified candidates are neither high-redshift nor BAL
quasars.

In both cases, the great majority of the low $P(Q)$ quasars are low
redshift objects: 24/34 (70\%) of the $P(Q)<0.2$ objects have
$z<0.25$.  These objects are probably being recognized as slightly
resolved on the POSS-I plates and so are, in some sense, rightly
classified as non-quasars (since they are perceptibly non-stellar.)

At the other extreme, the great majority of the high $P(Q)$ candidates
that are not quasars turn out to be BL~Lacs.  Using the classifier with
color, there are 419 objects with $P(Q)>0.85$, of which 373 (89\%) are
quasars and 46 (11\%) are non-quasars.  But 31 of the 46 non-quasars
are BL~Lacs.  If we combine the BL~Lacs with the quasars, an astounding
96.4\% of the high probability objects are quasars or BL~Lacs and only
3.6\% are other types of objects.

\placefigure{figpqfrac}

\preprint{
\begin{figure*}
\plotone{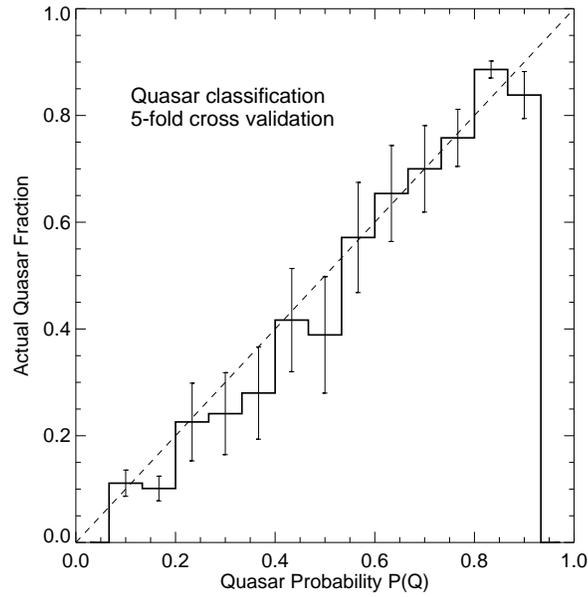}
\caption{
Fraction of candidates that are confirmed as quasars as a function
of $P(Q)$.  This was computed including color as one of
the parameters, but the results without color are essentially
identical.  The dotted line has a slope of unity.
The probability is a good predictor of the
likelihood that a candidate will turn out to be a quasar.
}
\label{figpqfrac}
\end{figure*}
}

Figure~\ref{figpqfrac} shows the fraction of candidates that turn out
to be quasars as a function of $P(Q)$.  The distribution follows well
the expected linear form assuming that $P(Q)$ is, in fact, a good
estimate of the probability that an object is a quasar.

\placefigure{figrc}

\preprint{
\begin{figure*}
\plotone{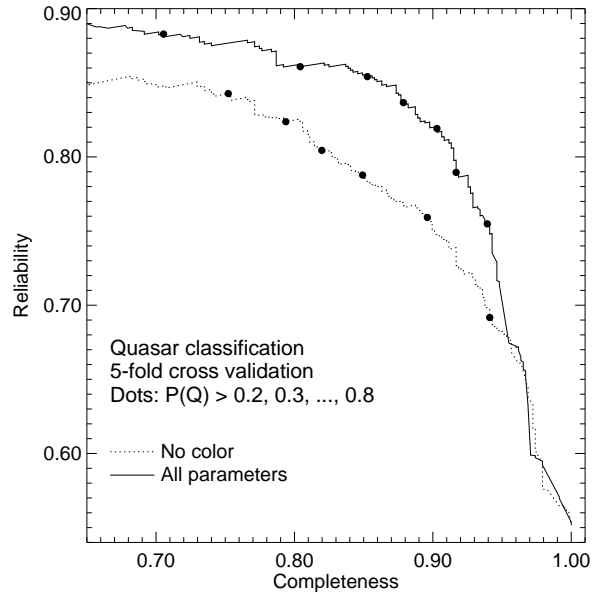}
\caption{
Completeness $C$ and reliability $R$ of the sample as a function of the
threshold selected for the probability $P(Q)$.  The completeness is the
fraction of quasars from the complete FBQS sample that are included at
the selected $P(Q)$ threshold; the reliability is the fraction of
selected candidates that are quasars.  The solid line is computed using
all parameters; the dashed line excludes color.  The FBQS sample
presented in this paper is the point at $C=1$, $R=0.55$ (i.e., the
sample is complete but only 55\% of the candidates turn out to be
quasars.) The efficiency of the quasar search (determined by the
reliability of the prediction) can be increased to $\sim89$\% using
color (85\% without color), if completeness is not crucial. Even with
completeness greater than 90\%, the reliability can be higher than
80\%.
}
\label{figrc}
\end{figure*}
}

Finally, Figure~\ref{figrc} shows how the completeness $C$ and
reliability $R$ of the sample vary as we change the $P(Q)$ threshold,
$P_C$, above which candidates are accepted into the sample.  If we set
$P_C=0$, all objects in the original FBQS sample are included, and the
resulting sample is 100\% complete (at least, it includes all the
objects in the current paper) but is only 55\% reliable (so 45\% of the
candidates turn out to be non-quasars.)  As $P_C$ increases, the
reliability of the sample increases dramatically at little cost in
completeness.  If color is used by the classifier, one can construct a
sample that is about 80\% reliable while still being 90\% complete, or
one can choose higher reliability (up to 89\%) if lower completeness
(70--80\%) is acceptable.  The results when color is not used are not
quite as good, but still it is possible to construct an impressively
pure sample at modest cost in completeness.

The way we might use this information depends on the scientific goals
of the quasar survey.  For the FBQS, our scientific interest is in
taking a census of quasars to understand their demographics and
physics.  Consequently, we put a high priority on completeness and are
willing to pay for it by spending extra telescope time to get spectra
of lower probability candidates.  We are loathe to risk losing high
redshift and heavily absorbed BAL quasars from our sample, so we prefer
to restrict the use of color and to aim instead for a highly complete
sample.

On the other hand, if one is interested in quasars mainly as sources to
probe the intergalactic medium, it could be appropriate to
sacrifice completeness and the occasional red quasar in order to get a
highly pure sample.  In a search for quasars at particular locations,
for example near other quasars or behind galaxy clusters, this approach
can be used to identify excellent candidates with high probability,
before a single night of telescope time has been used on spectroscopy.

To put our money where our mouths are, Table~8 gives our estimate
(using color parameters) of the probability $P(Q)$ that each object is
a quasar for the FBQS candidates that have not as yet been identified.
As these objects are observed, we are confident that our predictions
will be (in the mean) borne out.  While we expect to use our simple
sample-definition criteria as we continue expand the FBQS sample, we
are already using the $P(Q)$ probabilities to assist in prioritizing
targets for observation.  Further improvements are probably possible
utilizing advance knowledge of the X-ray and infrared properties of the
candidates.

\subsection{Summary of Incompleteness of the FBQS Sample}

In brief, the selection criteria applied to the FBQS sample lead to
several possible kinds of incompleteness and bias:

\begin{itemize}

\item The requirement of close positional coincidence excludes extended,
lobe-dominated radio sources with no core component.

\item The $E$ magnitude cut combined with the steep quasar luminosity
function leads to unavoidable Malmquist bias in the sample.  There is
also a radio Malmquist bias from the FIRST flux density limit, which
varies slightly as a function of position on the sky.

\item The requirement that the objects be stellar on the POSS-I plates
leads to incompleteness for low-redshift objects (which may appear
resolved) and will bias against the discovery of wide gravitational
lenses.  Since the APM classifier is not perfect, there can also be
truly stellar objects that are misclassified as galaxies on both plates
and so are missing from our sample.

\item The red color cut makes the sample incomplete for very high
redshift ($z\gtrsim3.5$) and heavily obscured quasars; however, most
such objects are likely to be optically too faint to be in the sample
anyway.

\item The FIRST catalog itself is incomplete for weak extended
sources.

\end{itemize}

\noindent
It is important to note that this survey is targeting quasars and hence
the selection criteria have been optimized for quasars. Generally
speaking, what works for quasars does not necessarily work for BL~Lac
objects, AGNs, or H~II galaxies. In particular, requiring the
optical morphology to be stellar and the optical colors to be bluer
than $O-E=2$ excludes very few quasars, but probably excludes the
majority of the other types of objects.  So even though we are finding
examples of such objects, the lists should not be taken as complete.

For most purposes, however, these selection effects do not seriously
affect the utility of the sample for statistical studies of quasars.
The biases in optically selected samples are generally worse in that they are
more difficult to characterize and correct.  For example, the differential
$K$-correction for broad-absorption line quasars introduces a complex
selection in number counts versus redshift.
The FBQS at least partly
redresses these biases (e.g., by allowing much redder objects to be
included.)  Indeed, the FBQS is, within its limits, perhaps the most
complete large-area quasar survey that exists.

\preprint{\bigskip}

\section{Optical Spectroscopy}
\label{sectionspectra}

Spectra for the quasar candidates are being collected at five different
observatories: the 3-m Shane telescope at Lick Observatory, the 2.1-m
telescope at Kitt Peak National Observatory\footnote{Kitt Peak National
Observatory, NOAO, is operated by the Association of Universities for
Research in Astronomy, Inc. (AURA), under cooperative agreement with
the National Science Foundation}, the 3.5-m telescope at Apache Point
Observatory\footnote{The Apache Point Observatory is owned and operated
by the Astrophysical Research Consortium.}, the $6\times1.8$-m
Multiple Mirror Telescope (``MMT classic''), and the 10-m Keck-II telescope.  The
spectrographs at the five observatories all have different resolutions
and wavelength coverage, which are summarized in Table~1. The observations
are made in a wide variety of atmospheric conditions ranging from
photometric to cloudy with both good and bad seeing. Many of the
candidates have been observed more than once. We have observed almost
all the candidates in the original candidate list, even those with
previous identifications.  (In several cases we have found errors in
published redshifts or positions.  Objects found to be discrepant
compared with the \cite{veron98} catalog are noted in the table.)

\placetable{table1}

\preprint{
\begin{deluxetable}{lcc}
\tablenum{1}
\tablewidth{0pt}
\tablecaption{Characteristics of Spectrographs\label{table1}}
\tablehead{
\colhead{} &
\colhead{Approximate} &
\colhead{} \\
\colhead{Telescope} &
\colhead{Wavelength Range (\AA)} &
\colhead{Resolution (\AA)}
}
\startdata
Lick 3-m     &       3600--8150     &        6 \\
KPNO 2.1-m   &       3700--7400     &        4 \\
APO 3.5-m    &       3650--10000    &       10 \\
MMT $6\times1.8$-m & 3600--8500     &        8 \\
Keck-II 10-m\tablenotemark{a} & 3800--8800 & 8 \\
\enddata
\tablenotetext{a}{\cite{oke95} describes the LRIS spectrograph on the
Keck-II telescope.}
\end{deluxetable}
}

\section{Spectroscopic Results}
\label{sectionresults}

Applying all the criteria described in \S\ref{sectionsample} results in
a list of 1238 candidates in 2682 square degrees, of which 169 are
previously known quasars.  In Tables 2--7 we present a list of all the
candidates in the FBQS sample with spectral classifications.  For each
object we list the FIRST catalog RA and Dec\footnote{The preferred
naming convention for objects in the sample is ``FBQS
Jhhmmss.s+ddmmss'', where the Right Ascension and Declination given in
the table are truncated (not rounded.)  This is consistent with the
IAU-approved naming convention for FIRST radio sources.} (J2000), the
recalibrated and extinction-corrected $E$ and $O$ magnitudes, the red
extinction correction $A(E)$, and the FIRST peak and integrated radio
flux densities.  The radio-optical positional separation and APM
star-galaxy classification (used to define the sample) are also given.
The objects have been segregated into 6 tables by their optical
spectral classification.  Quasars are listed in Table~2, narrow line
AGN in Table~3, BL~Lac objects in Table~4, H~II/star-forming
galaxies in Table~5, galaxies without any emission lines in Table~6,
and stars in Table~7. The criteria used to classify spectra are given
below.  Table~8 lists the objects for which spectra have not yet been
obtained.

For the purposes of the FBQS, we have classified any object with broad
emission lines as a quasar: i.e., we do not make a distinction between
quasars and Seyfert 1 galaxies.  We also make no distinction between
quasars and broad-line radio galaxies, the radio-loud counterparts to
Seyfert 1 galaxies.  The absolute blue magnitude is given for objects
with redshifts, so a conventional cut at $M_B = -23$ can be made if
desired to exclude lower-luminosity objects.  Of the 636 quasars, 50
fall into this lower luminosity category.  (The APM magnitudes are far
too bright for extended objects with apparent magnitudes brighter than
$\sim 12$~mag, so the high luminosities implied by the $M_B$ values for
such objects should not be taken seriously.  We have appended a note in
the table for such objects giving the blue magnitude from
NED\footnote{The NASA/IPAC Extragalactic Database (NED) is operated by
the Jet Propulsion Laboratory, California Institute of Technology,
under contract with the National Aeronautics and Space Administration.}
when available.)

BL~Lac classification is based on the traditional definition presented
in \cite{stocke91}:  all objects with emission line
rest-frame equivalent widths less
than 5~\AA\ and/or 4000~\AA\ break contrasts (Br$_{4000}$) less than
25\% appear in Table~4.  When emission lines or Br$_{4000}$ have been
measured, we list the quantities in the notes.  Although this
definition has been shown to be too restrictive, excluding some
objects that otherwise exhibit BL~Lac-like properties (see
\cite{marcha96} and \cite{laurent98}), our preliminary analysis
indicates that broader criteria admit many objects into the BL~Lac
class that do not properly belong there.  This is mainly because the
FBQS candidate objects are more than two orders of magnitude fainter in
the radio than the March\~{a} et al.\ sample used to define
the newer BL~Lac criteria. The FBQS selection thus admits many objects
with very low radio luminosities that exhibit weak optical emission and
absorption features like the high radio luminosity objects in
March\~{a} et al.  Many of these objects are clearly not BL~Lacs
despite being consistent with
the broader optical selection criteria.  A full
analysis of the BL~Lac content of the FBQS sample is underway
(\cite{laurent99}) and to avoid including non-BL~Lacs in the table, we
list only those objects that adhere to the strictest criteria.

We divide objects with narrow emission lines into narrow-line AGN and
H~II galaxies based on the ratios of [N~II] to H$\alpha$ and
[O~III] to H$\beta$.  An object is classified as a narrow-line AGN if
[N~II] is  more than 60\% of H$\alpha$, or (when H$\alpha$ is not
observed) if [O~III] is more than twice H$\beta$; otherwise it is
designated as an H~II galaxy (\cite{osterbrock89}).  The
classification based on [O~III]/H$\beta$ can be ambiguous and so the
H~II/AGN separation is less reliable for higher redshift
objects.

Some of the objects classified as H~II galaxies are
likely to be narrow-line Seyfert 1 galaxies, which can have similar line
ratios (\cite{osterbrock85}).  For this paper we have not attempted
to separate the two types of objects.

All other objects are classified as either normal galaxies (with no
obvious emission lines) or stars.  Galaxies flagged as previously known
in Table~6 have been identified in NED; in most cases we have not
obtained spectra, so their emission line (and Br$_{4000}$) status
should be considered less certain than the other objects in the table.

Note that most stars in the sample are likely to be chance coincidences
with the radio sources.  The APM positions for stars are degraded by
40~years of proper motion between the POSS-I and FIRST epochs; coupled
with the rarity of radio emission from stars, this prevents a useful
match between FIRST and APM for stars.  Recent epoch stellar catalogs
(ideally with proper motions) are required to study the radio emission
from stars using FIRST (\cite{helfandstars}).  One object listed in the
stars table, J163302.6+234928, is detected in the ROSAT All Sky Survey
Bright Source Catalog (\cite{voges96}) and is therefore likely to be a
true stellar radio source identification.  Another, J164018.1+384220,
is actually a known halo-population planetary nebula
(\cite{barker84}).  It is included with the stars because it is a
Galactic object (and it was formerly a star.)  A dozen or more of the
other objects in Table~7 are also likely to be real radio stars, but
current epoch optical positions will be needed to secure the
identifications.

In Tables 2--6, we list the measured redshift where available
(redshifts are unknown for many of the BL~Lac objects). Redshifts
for quasars and BL~Lac objects were computed by cross-correlating the
spectra with templates; for the quasars, a template constructed from
the FBQS objects themselves was used (\cite{brotherton99b}).
We use the
redshift to calculate for each object the radio luminosity
$L_R$ at a rest frequency of 5~GHz (assuming a radio spectral index of
$\alpha=-0.5$), the absolute $B$ magnitude $M_B$, and, as a measure of
radio loudness, the ratio $R^*$ of the 5~GHz radio flux density to the
2500~\AA\ optical flux in the quasar rest frame (assuming $\alpha_{opt}
= -1$ and using the definition of \cite{stocke92}).  We use the
(APS-calibrated) APM $O$ magnitude as a direct estimate of $B$, and we
do not correct the optical magnitude for the emission line
contribution.  The cosmological parameters
$H_0=50\,\hbox{km}\,\hbox{s}^{-1}\hbox{Mpc}^{-1}$,
$\Omega=1$, and $\Lambda=0$ were used in the luminosity
calculations.  When the redshift is not available we omit $M_B$ and
$L_R$, and we compute $R^*$ from the ratio of the 5~GHz flux density to
the 2500~\AA\ flux as above, but no $K$-corrections are applied.  There
is also a comments column, which notes details of particular interest
such as the presence of broad absorption lines or damped Lyman~$\alpha$
absorption lines, whether an object was previously known in the V\'eron
quasar catalog (\cite{veron98}), the Hamburg catalog (\cite{hagen99})
or in NED, association with
a ROSAT X-ray source or an IRAS infrared source, etc.

The spectra for all the objects identified as quasars are displayed in
Figure~\ref{figspectra} (See related series of gif files on the astro-ph
website).

%\preprint{\clearpage}

\section{Discussion}
\label{sectiondiscussion}

Detailed analysis of the FBQS sample will be deferred to other papers,
but we briefly discuss here both general properties of the sample and
some interesting source classes.

\placefigure{figzhist}

\preprint{
\begin{figure*}
\plotone{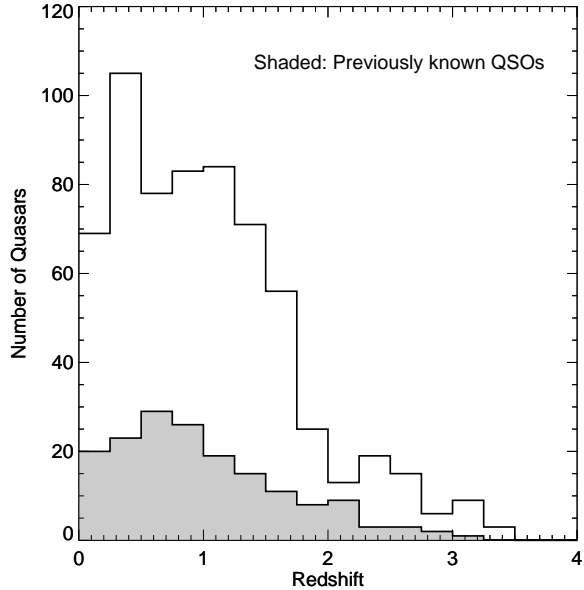}
\caption{
Histogram of redshifts for FBQS quasars.  Shaded: previously known quasars.
}
\label{figzhist}
\end{figure*}
}

\subsection{Redshift Distribution}

While the FIRST survey is clearly sensitive enough to detect quasars
out to redshifts greater than 3, the preponderance of FBQS quasars are
at redshifts below 2 (Fig.~\ref{figzhist}).  The precipitous drop
beyond $z=2$ is attributable more to the limit in $E$ than to the limit
in radio flux density.  It is apparent in the redshift-radio luminosity
distribution (Fig.~\ref{figlumz}a) that the high-redshift quasars do
not crowd against the radio detection limit; they are missing from the
sample mainly because they are too faint in the optical
(Fig.~\ref{figlumz}b).  On the other hand, the lack of quasars at
$z>3.5$ could be partly due to the red color cut used to define the
sample, though the number of quasars beyond $z=3$ that are lost due to
the color limit is certainly small (\S\ref{sectioncolorcut}).

\placefigure{figdetfrac}

\preprint{
\begin{figure*}
\plotone{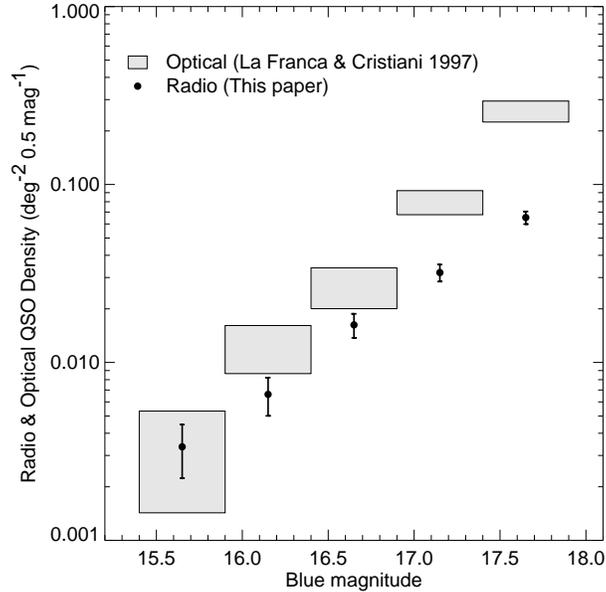}
\caption{
Density of FBQS quasars on the sky as a function of
extinction-corrected blue magnitude.  Shaded boxes show the density of
all optically selected quasars from La Franca \& Cristiani (1997).  The
redshift range of FBQS quasars is restricted to $0.3\le z \le 2.2$ to
match the optical counts.  The surface density of FBQS quasars is very
similar to optically selected samples for $B<16$~mag, indicating that
FIRST detects most optical quasars and that the FBQS and optical
samples are similarly incomplete for bright objects.  At $B=18$~mag the
FBQS sky density drops to $\sim25$\% of the optically selected quasar
density.
}
\label{figdetfrac}
\end{figure*}
}

\subsection{Efficiency of the FBQS}

The absolute efficiency of the FBQS in identifying quasars can be
estimated by comparing the number of quasars found in the FBQS to the
number expected from previous surveys. Figure~\ref{figdetfrac} compares
the sky density of FBQS quasars to estimates of the sky density of
optically selected quasars as given by \cite{lafranca97}.  A correction
factor of 1.17 has been applied to the FBQS counts in the
$B=17.4$--17.9 bin to account for objects that do not appear in the
FBQS either because they are too faint on the red plate ($E > 17.8$) or
because they are not yet identified (Table~8).  The correction factor
for the other bins is negligible, changing the raw counts by less than
3\%.

For quasars brighter than $B=16.4$~mag, the FBQS quasar density is
indistinguishable from the density of optically selected quasars, so
the efficiency is very high.  The efficiency drops at fainter
magnitudes, falling to $\sim25$\% when $B=17.4$--17.9~mag.  This does
not mean that FIRST detects all optically bright quasars --- we know
that there are radio-silent quasars that would not be detected by
FIRST.  We can conclude, however, that FIRST detects the bulk of bright
radio-quiet quasars and that the incompleteness of the FBQS (due mainly
to very low radio fluxes) is similar to the incompleteness of bright
optically selected samples (which is typically due to objects missed due to
unusual color and emission line properties.)

An analysis of both quasars in the \cite{veron98} and Hamburg (\cite{hagen99})
catalogs indicates that approximately 25\% of all quasars brighter than
$E=17.8$ are detected by the FIRST survey.  The detected fraction is a bit
lower for the Hamburg catalog -- FIRST detects 17 out of 78 (22\%) quasars in the
FBQS area.  That may just be a reflection of the statistics of the small numbers
involved, or it may indicate that the Hamburg survey is more efficient than
previous optical surveys at detecting quasars with unusual optical properties.

\subsection{Advantages Over Optical Surveys}

When compared to previous quasar surveys, we find that the FBQS has two
primary advantages.  First, because it is a radio-selected survey, it
has high selection efficiency, mainly because detectable radio emission
from stars is extremely rare.  Only about one in $10^4$ 12th magnitude
stars has radio emission detectable by the FIRST survey
(\cite{helfandstars}).  In contrast, X-ray selected samples are
contaminated by ubiquitous stellar X-ray emission.  The FBQS can cover
a significantly larger sky area than optical objective prism surveys,
with very much less telescope time required than for optical
color-selected surveys.  Perhaps the color-selection method will become
comparably efficient with the advent of the wide-area, 5-color Sloan
Digital Sky Survey, but until then radio selection is the most
efficient technique for discovering new quasars (once someone has done
the hard work of making a deep, wide-area radio catalog.)

Second, the radio-selected FBQS can discover class\-es of quasars that
are rare or unknown in other surveys because of biases against them.
The best example in the FBQS sample are the extreme broad absorption
line quasars that show low ionization iron absorption.  \cite{becker97}
reported the discovery of two such objects, J084044.4+363328 ($z=1.22$;
included in this paper) and J155633.8+351758 ($z=1.48$; too faint for
the FBQS.)  This paper includes 3 additional objects: J104459.5+365605
($z=0.701$), J121442.3+280329 ($z=0.700$), and J142703.6+270940
($z=1.170$).  The spectra of these objects are remarkable; they are
often difficult even to recognize as quasars.  The emission lines are
almost completely masked by absorption, and the colors of these objects
are quite red.  Such objects would be very unlikely to be recognized in
any existing optically selected survey, so we really have no idea how
common they may be among radio-quiet quasars.

In total there are 29 definite or tentative BAL quasars among the 636
FBQS quasars.  This is comparable to the percentage of BAL quasars in
optically selected samples (\cite{foltz90}).  The properties of the
FBQS BAL quasar sample are discussed further by \cite{becker99}.

Of course, the disadvantage of a radio-selected survey is that we can
find only quasars with radio emission above the FIRST 1~mJy limit, but
as discussed above
these constitute a significant minority (25\%) of all quasars
brighter than $E=17.8$.
The FIRST survey is sufficiently deep that we detect many
radio-intermediate and even radio-quiet quasars; only radio-silent
quasars are not represented in the sample.  This is most apparent in
the number counts (Fig.~\ref{figdetfrac}), which are very similar for
FBQS and optically selected samples for the brightest objects.  It is
the combination of large optical quasar surveys with the FBQS that may
be most productive for studying quasar physics.

\subsection{No Radio-Loud/Radio-Quiet Dichotomy?}
\label{sectionbimodal}

\placefigure{figrz}

\preprint{
\begin{figure*}
\plotone{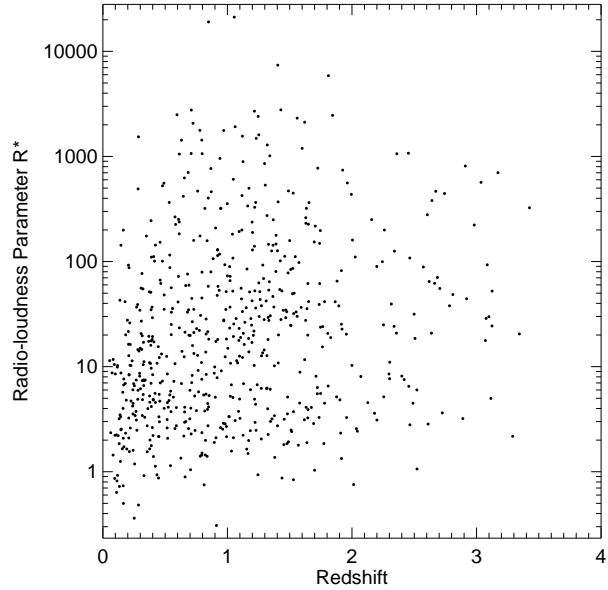}
\caption{
Radio-optical ratio $R^*$ (rest-frame ratio of the 5~GHz radio flux
density
to the 2500~\AA\ optical flux; \cite{stocke92}) versus redshift for
FBQS quasars.
}
\label{figrz}
\end{figure*}
}

Figure~\ref{figrz} shows the distribution of the radio-loudness
parameter $R^*$ versus redshift.  An $R^*$ of 10 is generally taken as
the divide between radio-loud and radio-quiet objects (\cite{stocke92}).  It is
clear that over the entire range of observed redshifts, the FBQS is
reaching well into the radio-quiet population and that there is no obvious
deficit of quasars at the boundary.  Previous studies, in contrast,
have always found a bimodal distribution in $R^*$ with a dearth of
quasars in the range $3\lesssim R^* \lesssim 100$ (e.g.,
\cite{miller90}; \cite{stocke92}).

\placefigure{figrhist}

\preprint{
\begin{figure*}
\begin{tabular}{cc}
\epsfxsize=0.45\textwidth \epsfbox{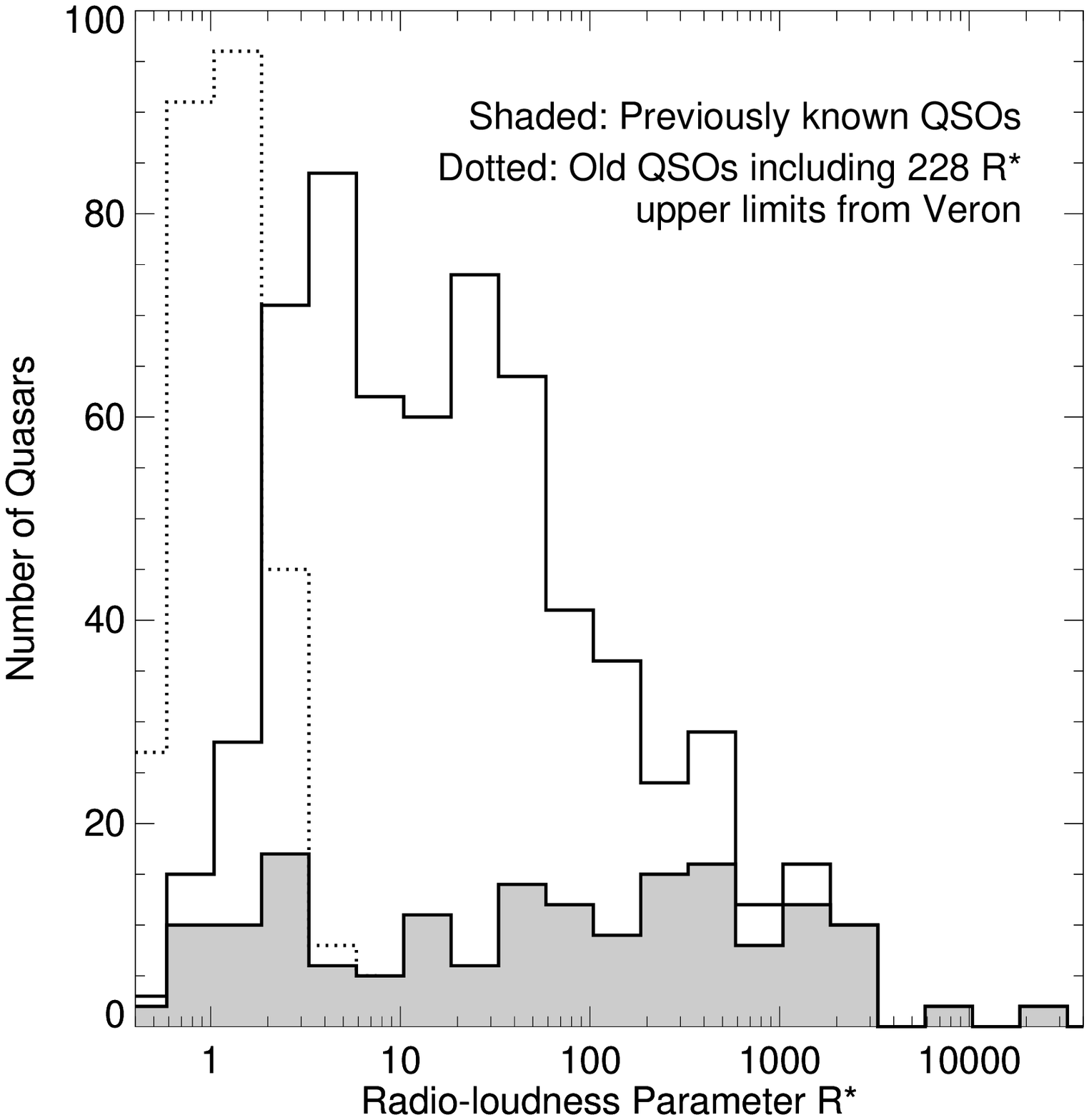} &
\epsfxsize=0.45\textwidth \epsfbox{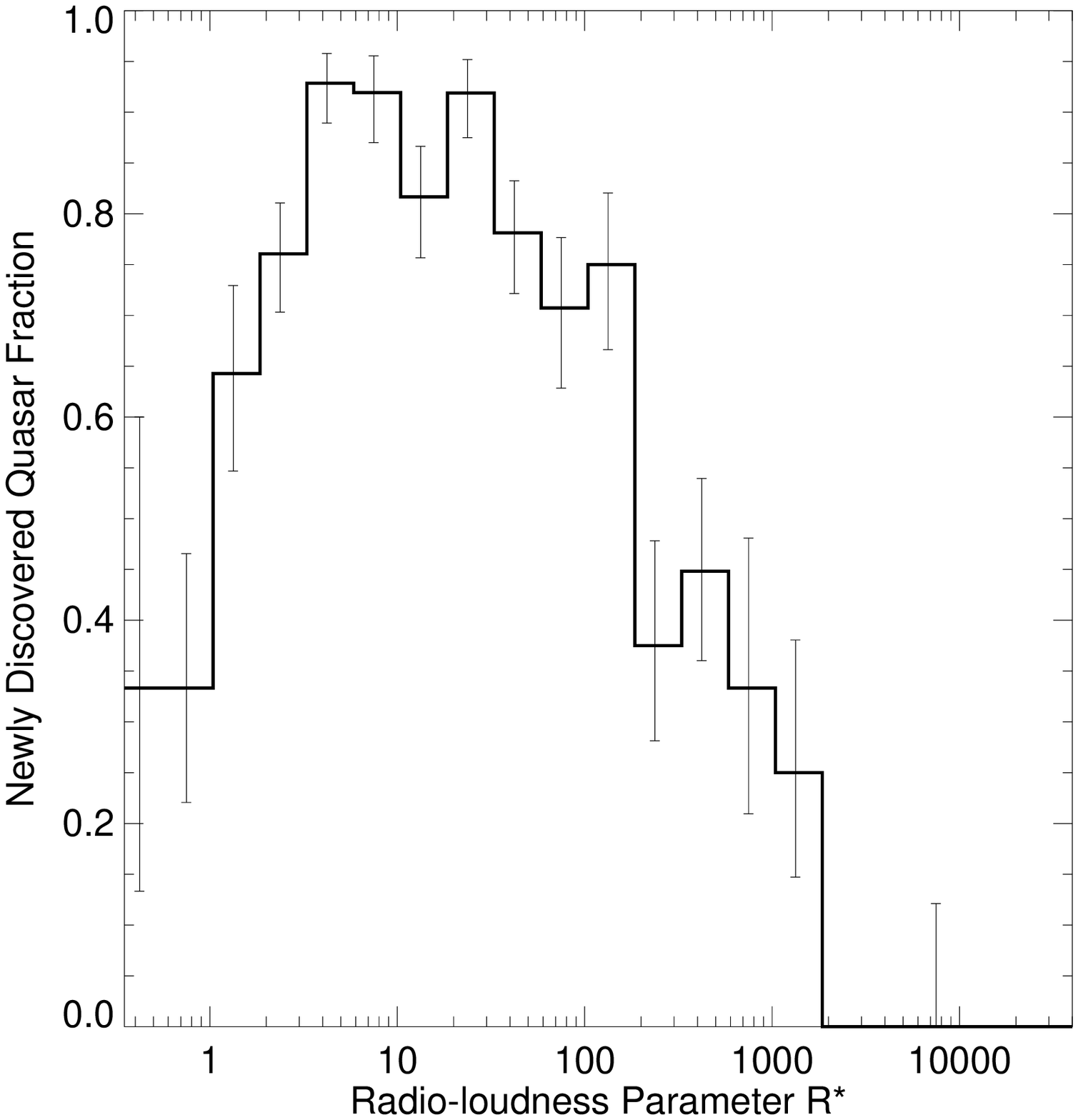} \\
(a) & (b) \\
\end{tabular}
\caption{
(a) Histogram of radio-optical ratio $R^*$ (\cite{stocke92}) for FBQS
quasars.  Shaded: previously known quasars.  The dotted histogram
includes $R^*$ upper limits for V\'eron catalog objects in the FBQS area but
not detected by FIRST.  The V\'eron quasars (shaded plus dotted histograms)
show a bimodal distribution
of $R^*$, with a dip around $R^* = 3$--30, but the FBQS quasar counts
rise continuously through that region and show no obviously evidence
for bimodality.  (b) Fraction of quasars that were newly discovered
versus $R^*$.  The FBQS is increasing the number of known objects in
the radio-quiet/radio-loud transition region ($R^*=1$--100) by a large
factor.
}
\label{figrhist}
\end{figure*}
}

The inadequacy of previous surveys in finding these radio-transition
quasars is well illustrated in Figures~\ref{figrhist}(a) and
\ref{figrhist}(b), which show the histogram of the number of quasars
versus $R^*$ compared with the distribution for previously known
quasars. Although the FBQS shows a sharp increase in the number of
quasars for $R^* < 100$, this population is actually decreasing among
the ranks of previously known quasars.  The vast majority (85\%) of
FBQS quasars with $1<R^*<100$ are newly discovered, while those with
more extreme $R^*$ values in both directions are well-represented in
existing samples.

Figure~\ref{figrhist}(a) also shows a histogram for previously known
quasars that includes $R^*$ upper limits for all the V\'eron quasars
(\cite{veron98}) that fit the FBQS criteria (including sky area and
brightness) but that are not detected by the FIRST survey.  This
distribution is certainly bimodal; however, the FBQS quasars fill the
gap and there is no clear evidence for bimodality in the combined FBQS
and radio-quiet V\'eron distribution.

Conceivably this is the result of selection effects in the FBQS, though
we have no plausible explanation for how our sample selection could
eliminate a gap in the $R^*$ distribution.  A definitive conclusion on
the actual distribution of quasar radio-loudness awaits full analysis
and modeling of the sample and its selection effects.

\subsection{Summary}

This paper purposely defers most of the scientific studies that
can be based on this new sample of quasars in order to make the data
available to the community more quickly. Clearly the value of this
database will be greatly enhanced with complementary data at other
radio frequencies as well as in the X-ray, IR, and UV bands.  The FBQS
will serve as a vital bridge between the traditional radio-loud and
radio-quiet quasar samples. The literature is full of characteristics
such as line ratios, X-ray spectral indices, and broad absorption lines
that appear to depend on radio-loudness. With the new sample, we can begin
to explore the radio-loudness transition zone.

The results in this paper are based on the 1997 version of the FIRST
radio catalog, which covered 2682 square degrees of the north Galactic
cap.  As this is written in mid-1999, data have been collected
that extend the area covered by FIRST to 6000~square degrees, and the
FBQS candidate list will soon stand at $\sim2700$; $\sim1500$ of those
candidates will eventually be confirmed as quasars (assuming that the
telescope time allocation committees continue to be kind to us.)  At
that point the FBQS catalog will be one of the largest uniformly
selected quasar surveys and will be an even more powerful tool for
furthering our understanding of the quasar phenomenon.

\acknowledgments

Thanks to the referee, Todd Tripp, for many helpful suggestions, including
an analysis of the FBQS/Hamburg match rate.
We appreciate helpful discussions with Ed Moran on the classification
of the emission line galaxies.  Thanks to Hien Tran for computing the
quasar redshifts from template correlations.  We acknowledge extensive use of the
NASA/IPAC Extragalactic Database (NED), which is operated by the Jet
Propulsion Laboratory, Caltech, under contract with the National
Aeronautics and Space Administration.  The success of the FIRST survey
is in large measure due to the generous support of a number of
organizations.  In particular, we acknowledge support from the NRAO,
the NSF (grants AST-98-02791 and AST-98-02732), the Institute of
Geophysics and Planetary Physics (operated under the auspices of the
U.S. Department of Energy by Lawrence Livermore National Laboratory
under contract No.~W-7405-Eng-48), the Space Telescope Science
Institute, NATO, the National Geographic Society (grant NGS
No.~5393-094), Columbia University, and Sun Microsystems.  We
appreciate the hospitality of the Institute of Astronomy,
Cambridge University, which hosted visits
enabling the writing of much of this paper.

%\placefigure{figspectra}

\clearpage

\preprint{

% Tables

\begingroup
\tabcolsep=3pt
\footnotesize

\endgroup

% figure with spectra

\clearpage

\begin{figure*}
\caption{
Spectra of FBQS candidates identified as quasars, sorted by
decreasing redshift.
The dotted lines show expected
positions of prominent emission lines:
Ly$\alpha$~1216,
N~V~1240,
Si~IV~1400,
C~IV~1550,
C~III]~1909,
Mg~II~2800,
H$\delta$~4102,
H$\gamma$~4341,
H$\beta$~4862,
[O~III]~4959,
[O~III]~5007,
H$\alpha$~6563.
Note that most of the spectra have atmospheric
A and B band absorption at $\sim6880$~\AA\ and 7620~\AA.
PLEASE SEE THE RELATED GIF FILES FOR THE ACTUAL SPECTRA.
}
\label{figspectra}
\end{figure*}
\clearpage

\end{document}